\documentclass[notitlepage,letterpaper]{article}
\usepackage[ansinew]{inputenc} 
\usepackage{amsmath}
\usepackage{amsfonts}
\usepackage{amssymb}
\usepackage[colorlinks=true,urlcolor=blue,linkcolor=blue]{hyperref} 
\usepackage{graphicx}
\usepackage{geometry}      
\usepackage{epstopdf}
\usepackage{fancyhdr} 

\begin{document}
\title{ \textbf{Variable Eddington Factor and \\ Radiating Slowly Rotating Bodies \\ in
General Relativity}}
\author{
\textbf{ F. Aguirre }\thanks{e-mail: \texttt{aguirre@ula.ve }
\qquad Web: \url{http://webdelprofesor.ula.ve/ciencias/aguirre/}} \\
\textit{ Laboratorio de Física Teórica, Departamento de Física, Facultad de Ciencias} \\\textit{Universidad de Los Andes, Mérida 5101, Venezuela } \\
\textbf{L. A. Núñez}\thanks{e-mail: \texttt{nunez@ula.ve }\qquad Web:
\url{http://webdelprofesor.ula.ve/ciencias/nunez/}} \\ 
\textit{Centro de Física Fundamental,}  
\textit{Departamento de Física, Facultad de Ciencias, }\\
\textit{Universidad de Los Andes, Mérida 5101, Venezuela} and \\ 
\textit{Centro Nacional de Cálculo Científico, Universidad de Los Andes,}\\ 
\textsc{(CeCalCULA),}  \\
\textit{Corporación Parque Tecnológico de Mérida, Mérida 5101, Venezuela,} and \\
\textbf{T. Soldovieri}\thanks{e-mail: \texttt{tsoldovieri@luz.edu.ve}} \\
\textit{Laboratorio de Astronomía y Física Teórica (LAFT),} \\
\textit{Departamento de Física, Facultad de Experimental de
Ciencias, }\\ \textit{Universidad del Zulia, Maracaibo Zulia, Venezuela}
}
\date{Version 2.0 July 2006 \\
PACS 04.40.Dg 95.30.Jx 95.30.Sf 97.60.Jd }
\maketitle

\begin{abstract}
We present an extension to a previous work to study the collapse of a radiating, slow-rotating self-gravitating relativistic configuration. In order to simulate dissipation effects due to the transfer of photons and/or neutrinos within the matter configuration, we introduce the flux factor, the variable Eddington factor and a closure relation between them. Rotation in General Relativity is considered in the slow rotation approximation, i.e. tangential velocity of every fluid element is much less than the speed of light and the centrifugal forces are little compared with the gravitational ones. Solutions are properly matched, up to the first order in the Kerr parameter, to the exterior Kerr-Vaidya metric and the evolution of the physical variables are obtained inside the matter configuration. To illustrate the method we explore the influence of the closure relations on the dynamics of three models with different equations of state and two functional form of the flux factor. We have found that, for the six closure relations considered, the matching conditions implies that a total diffusion regime can not be attained at the surface of the configuration. It has also been obtained that the eccentricity at the surface of radiating configurations is greater for models near the diffusion approximation than for those in the free streaming out limit. At least for the static ``seed'' equations of state considered, the simulations we performed show that these models have differential rotation and that the more diffusive the model is, the slower it rotates.
\end{abstract}

\section{Introduction}

Compact objects are one of the most fascinating objects known in our Universe. White dwarfs, neutron stars, quark stars, hyperon stars, hybrid stars and magnetars are thought to be relics from most of the cores of luminous stars which we believe to be born in supernova explosions. 

Core collapses are triggered by the implosion of the inner nucleus of a massive star ($M_{\star} \sim8-20M_{\odot}$) when its mass is in the limit of Chandrashekar ($M_{core}\sim1.4M_{\odot}$). During the implosion
nearly all of an enormous gravitational binding energy ($ ({GM^{2}} )/{R} \sim 5\times10^{53} $ ergs $\sim0.2Mc^{2}$) gained is stored as internal energy of a newly born, proto-neutron star (PNS) and driven by neutrino diffusion which cools this new type of compact object. Temporal and spectral characteristics of the neutrino emission depend on the rate at which they diffuse through the imploded PNS which, at this early stage, would have a mean density several times the standard nuclear density, $\bar{\rho}\sim3M/ \left( 4\pi R^{3}\right)  $ $\approx7\times10^{14}g\ cm^{-3},$ with $\rho_{0} \simeq 2.\times10^{14}g\ cm^{-3}$. The core density reaches up to $(10-20)\rho_{0}$ during the cooling time $t_{cool}\sim5$ to $10$s, while the PNS de-leptonizes by neutrino emission, cools,
contracts and spin-ups to form the final ultradense compact object \cite{Demianski1985,ShapiroTeukolsky1983,KippenhahnWeigert1990,Glendening2000}.

There is a consensus that the above standard scenario requires the description of General Relativity because of the formidable gravitational fields arising during these processes. These powerful gravitational fields strongly couple hydrodynamics and neutrino flows within rotating matter configurations (see \cite{BruennDeNiscoMezzacappa2001} for a good historical survey of previous works done at various levels on the problem of coupling the General Relativity, hydrodynamics, and radiation transport in spherical symmetry). Unfortunately despite a considerable effort that is been carried out by a significant group of people and institutions, presently we do not have any self-consistent model either analytical or numerical that includes all of those components in full details.

Although an exterior metric of a rapidly rotating neutron agrees with the
corresponding Kerr metric only to lowest order in the rotational velocity  \cite{HartleThorne1969}, there have been many attempts to find a closed interior solution which matches smoothly to the gravitational
field outside a rotating source (see \cite{Stergioulas2003,Font2003,Lorimer2001} and references therein). In general these attempts have proved to be unsuccessful essentially because the considerable mathematical complexity in solving the Einstein equations \cite{ChineaGonzalez-Romero1993}. It is only very recently, that there has been reported some progress in the\ analytical approach \cite{MankoMielkeSanabria2000} which approximately match numerical solutions for rapidly rotating neutron stars \cite{BertiStergioulas2004,BertiEtal2004,PachonRuedaSanabria2006} and has been used in studies of energy release
 \cite{SibgatullinSunyaev2000,SiebelEtall2003}.\ Recent numerical research has also considerably advanced our understanding of rotating relativistic stars \cite{DimmelmeierFontMueller2002}. There now exist several independent numerical codes for obtaining accurate models of rotating neutron stars in full General Relativity (see \cite{Stergioulas2003} and \cite{Font2003} for a good review on this subject). We can particularly mention a 3D
general-relativistic hydrodynamics code (GR Astro) \cite{NASAGC,GRAstro3D} and built from the Cactus Computational Toolkit \cite{Cactus} and some recent contribution for supernova scenario  incorporating a better microphysics into rotating modeling of compact objects \cite{WalderEtal2005}.

Simulations which include better microphysics in the form of realistic nuclear equations of state or neutrino transport have either been confined to spherical symmetry or restricted to newtonian gravity. Today, all available models analytical/numerical resembling some pieces of truth, demonstrate remarkable sensitivities to different physical aspects of the problem, in particular the treatment of neutrino transport and neutrino-matter interactions, the properties of the nuclear equation of state (EoS), multi-dimensional hydrodynamical processes, effects of rotation and general relativity. It is worth mentioning that is just recently, when it has become possible to obtain spherically symmetric general relativistic hydrodynamical core-collapse, treating the time and energy dependent neutrino transport in hydrodynamical simulations by considering a Boltzmann solver for the neutrino transport
\cite{LiebendorferEtal2001,LiebendorferEtal2002}, implementing multigroup flux-limited diffusion to Lagrangian Relativistic Hydrodynamics \cite{BruennDeNiscoMezzacappa2001} or assuming the variable Eddington factor method to deal with the integro-differential character of the Boltzmann equation \cite{RamppJanka2002}.

The present paper lies in between the traditional analytical and the emerging numerical descriptions of gravitational collapse. It follows a seminumerical approach which considers, under some general and reasonable physical assumptions, the evolution of a general relativistic rotating and radiating matter configurations. The rationale\ behind this work is twofold, first it seems useful to consider relatively simple nonstatic models to analyze some essential features of realistic situations that purely numerical solutions could hinder. Particularly, we will focus on the influence upon the evolution of matter configurations of the dissipation mechanism due to the emission of photons/neutrinos. Secondly it could be helpful for the evolving numerical codes to have testbed arena including General Relativity, rotation, dissipation and plausible EoS.

The approach we follow to solve the Einstein Equations starts from heuristic assumptions relating density, pressure, radial matter velocity and choosing a known interior (analytical) static spherically symmetric ( considered as ``seed'') solution to the Tolman-Oppenheimer-Volkov equation. This scheme transforms the Einstein partial differential equations into a system of ordinary differential equations for quantities evaluated at surfaces whose
numerical solution, allows the modelling of the dynamics of the configuration. This method is an extension of the so called HJR  \cite{HerreraJimenezRuggeri1980}, which has been successfully applied to a variety of astrophysical scenarios (see \cite{HerreraNunez1990} and \cite{HernandezNunezPercoco1998} and references therein) and which has been recently revisited \cite{BarretoMartinezRodriguez2002,HerreraEtal2002} in order to
appreciate its intrinsic worth.

We are going to consider the effects of rotation in General Relativity in the slow rotation approximation, i.e. up to the first order, thus we shall maintain only linear terms in the angular velocity of the local inertial frames. Thus, the effects of rotation are purely relativistic and manifest through the dragging of local inertial frames
\cite{Hartle1967,HartleThorne1968}. This is understandable if we recall that in the newtonian theory where the parameter measuring the ``strength'' of rotation (the ratio of centrifugal acceleration to gravity at the equator) is not linear in the angular velocity but proportional to the square of it. The slow rotation approximation has recently proved to be very reliable for most astrophysical applications \cite{BertiEtal2004}. This assumption is very sensible because it considers that the tangential velocity of every fluid element is much less than the speed of light and the centrifugal forces are little compared with the gravitational ones. It is worth mentioning that the continuity
of the \textit{first} and the \textit{second fundamental} ($g_{ij}$ and $K_{ij}$) forms across the matching surface are also fulfilled up to this order of approximation.

Conscious of the difficulties to cope with dissipation due to the emission of photons and/or neutrinos and, aware of the uncertainties of the microphysics when considering the interaction between radiation and ultradense matter, we extend a previous work \cite{HerreraEtal1994} to study the collapse of a radiating, slow-rotating self-gravitating relativistic configuration by introducing a relation between the radiation energy flux density and the radiation energy density, i.e. the \textit{flux factor,} $f = {\mathcal{F}}/{\rho_{R}},$ and the so called \textit{variable Eddington factor,} $\chi= {\mathcal{P}}/{\rho_{R}}, $ relating the radiation pressure and the radiation energy density, We have also include a closure relation between both quantities, i.e., $\chi=\chi(f).$ In the literature several closures have been introduced (see \cite{Levermore1984} for a comprehensive review and \cite{Dominguez1997}, \cite{PonsIbanezMiralles2000} and \cite{SmitVandenHornBludman2000} for more recent references) and most of them are consistent with the hiperbolicity and causality required by a relativistic theory \cite{PonsIbanezMiralles2000}.

With the above set of assumptions, i.e. seminumerical approach to
solve the Einstein system, slow rotation approximation, and a
particular closure relation between the flux and the variable
Eddington factor, we shall explore the effect of dissipation on
the evolution of the rotating radiating matter configuration. The
outcomes from our simulations could represent rotating compact
objects where the core rotates faster than the envelope.
Therefore, the core can be supported by rapid rotation while the
velocity of the fluid at the equator does not exceed the limit
imposed by a fluid moving along a geodesic (the Kepler limit).
Thus, differential rotation may play an important role for the
stability of these remnants, since it can be very effective in
increasing their maximum allowed mass. This effect was
demonstrated in newtonian gravitation in
 \cite{OstrikerBodenheimerLyndenBell1966} and was recently found
by Shapiro and collaborators for general relativistic
configurations having a polytropic EoS
 \cite{BaumgartShapiroShibata2000,LyfordBaumgarteShapiro2002}. For
these rotating matter distribution we have found that boundary
conditions imply that at the surface of the configuration, a total
diffusion regime can not be attained. It has also been obtained
that, with these coordinates, the eccentricity at the surface of
radiating configurations, (up to first order) is greater for
models near the diffusion limit approximation than for those in
the free streaming out limit even though the rotation of
configurations with dissipation near the diffusion limit appears
to be slower than those near the free streaming out limit. We
noticed that, at least for the static ``seed'' equations of state
considered, it seems that the lower the flux factor we have, the
slower is the rotation of the configuration.

The plan for the present work is the following. The next section contains an
outline of the general conventions, notation used, the metric, the structure
of the energy tensor and the corresponding field equations. Section
\ref{eddingtonfactor} is devoted to describe the variable Eddington factor,
the closure relations and the limits for the radiation field. Junction
conditions and their consequences are considered in Section
\ref{JunctionConditions}. The method is sketched in Section \ref{TheMethod}.
We work out the modelling, previously studied for the spherical (nonrotating)
case in Section \ref{Modeling}. Finally some comments and conclusions are
included in \ref{Conclusions}.

\section{Energy-momentum tensor and field equations}

\subsection{The metric}

As in the previous work \cite{HerreraEtal1994}, let us consider a
nonstatic, axially symmetric distribution of matter conformed by
fluid and radiation where the exterior metric, in radiation
coordinates\cite{Bondi1964}, is the Kerr-Vaidya
metric\cite{CarmeliCaye1977}:
\begin{align}
ds^{2} &  =\left(  1-\frac{2m\left(  u\right)  \ r}{r^{2}+\alpha^{2}\cos
^{2}{\theta}}\right)  \mathrm{d}u^{2}+2\mathrm{d}u\mathrm{d}r-2\alpha\sin
^{2}{\theta}\mathrm{d}r\mathrm{d}\phi+4\alpha\sin\theta^{2}\frac{m\left(
u\right)  \ r}{r^{2}+\alpha^{2}\cos^{2}{\theta}}\mathrm{d}u\mathrm{d}%
\phi\nonumber\\
& \label{mexterna}\\
&  \qquad\qquad\qquad-(r^{2}+\alpha^{2}\cos^{2}{\theta})\mathrm{d}\theta
^{2}-\sin^{2}{\theta}\left[  r^{2}+\alpha^{2}+\frac{2m\left(  u\right)
\ r\ \alpha^{2}\sin^{2}{\theta}}{r^{2}+\alpha^{2}\cos^{2}{\theta}}\right]
\mathrm{d}\phi^{2}\;.\nonumber
\end{align}
Here, $m\left(  u\right)  $ is the total mass and $\alpha$ is the
Kerr parameter representing angular momentum per unit mass in the
weak field limit. It is worth mentioning at this point that the
metric above is not a pure radiation solution and may be
interpreted as such only asymptotically
\cite{GonzalezHerreraJimenez1979}. A pure rotating radiation
solution may be found in reference \cite{KramerHahner1995}.
However, as we shall show below although the interpretation of the
Carmeli-Kaye metric in not completely clear, the model dependence
of the considered effect is independent of the shape and the
intensity of the emission pulse, and may be put in evidence even
for a tiny radiated energy, $\Delta M_{rad}=10^{-12}M(0)$, which
for any practical purpose corresponds to the Kerr
metric\cite{HerreraEtal1994}. The interior metric is written as
\cite{HerreraJimenez1982}
\begin{align}
ds^{2} &  =e^{2\beta}\left\{  \frac{V}{r}\mathrm{d}u^{2}+2\mathrm{d}%
u\mathrm{d}r\right\}  -(r^{2}+\tilde{\alpha}^{2}\cos^{2}{\theta}%
)\mathrm{d}\theta^{2}+2\tilde{\alpha}e^{2\beta}\sin^{2}{\theta}\left\{
1-\frac{V}{r}\right\}  \mathrm{d}u\mathrm{d}\phi\nonumber\\
& \label{minterna}\\
&  \qquad\qquad\qquad-2e^{2\beta}\tilde{\alpha}\sin^{2}{\theta}\mathrm{d}%
r\mathrm{d}\phi-\sin^{2}{\theta}\left\{  r^{2}+\tilde{\alpha}^{2}%
+2\tilde{\alpha}^{2}\sin^{2}{\theta}\frac{V}{r}\right\}  \mathrm{d}\phi
^{2}.\nonumber
\end{align}

In the above equations (\ref{mexterna}) and (\ref{minterna}), $u~=~x^{0}$ is a
time like coordinate, $r~=~x^{1}$ is the null coordinate and $\theta~=~x^{2}$
and $\phi~=~x^{3}$ are the usual angle coordinates. Local minkowskian
coordinates $(t,x,y,z)$ are related to Bondi radiation coordinates
$(u,r,\theta,\phi)$ by%

\begin{equation}
\mathrm{d}t   =e^{\beta}\left(  \sqrt{\frac{V}{r}}\mathrm{d}u+\sqrt{\frac{r}{V}}\mathrm{d}r\right)  + e^{\beta}\widetilde{\alpha}\sin^{2}\theta \left(  \sqrt{\frac{r}{V}}-\sqrt{\frac{V}{r}}\right)  \mathrm{d}\phi;\label{minKdt}
\end{equation}
\begin{equation}
\mathrm{d}x   =e^{\beta}\sqrt{\frac{r}{V}}\left(  \mathrm{d}r+\widetilde{\alpha}\sin^{2} \theta  \mathrm{d}\phi\right)  ;\quad
\mathrm{d}y=\sqrt{r^{2}+\widetilde{\alpha}\cos^{2} \theta} \mathrm{d}\theta \quad
\text{and} \quad \mathrm{d}z=\sin \theta  \sqrt{r^{2}+\widetilde{\alpha}\cos^{2} \theta  }\mathrm{d}\phi.
\label{minKdxdydz}
\end{equation}
The $u$-coordinate is the retarded time in flat space-time, therefore,
$u$-constant surfaces are null cones open to the future. This last fact can be
readily noticed from the relationships between the usual Schwarzschild
coordinates, $(T,R,\Theta,\Phi)$, and Bondi's radiation coordinates:
\begin{equation}
u=T-\int\frac{r}{V}\;dr,\hspace{0.5cm}\theta=\Theta,\hspace{0.5cm}%
r=R\hspace{0.5cm}\text{and}\hspace{0.5cm}\phi=\Phi\;,\label{eq_schcoord}%
\end{equation}
which are valid, at least, on the surface of the configuration.

The Kerr parameter for the interior space-time (\ref{minterna}) is denoted
$\tilde{\alpha}$ and, for the present work it is relevant only (as well as
$\alpha$ in eq. (\ref{mexterna})) up to the \textit{first order}. Notice that,
in these coordinates, the $r~=~r_{s}~=~const,$ represent surfaces that are not
spheres but\textit{\ oblate spheroids}, whose eccentricity depends upon the
interior Kerr parameter $\tilde{\alpha}$ and is given by
\begin{equation}
e^{2}=1-\frac{r_{s}^{2}}{r_{s}^{2}+\tilde{\alpha}^{2}}\;. \label{eccentricity}%
\end{equation}
Observe that this expression of eccentricity yielding the correct newtonian
limit and corresponding to the natural definition in the context of metrics
(\ref{mexterna}) and (\ref{minterna}), is not invariantly defined .

The metric elements $\beta$ and $V$ in eq. (\ref{minterna}), are functions of
$u$, $r$ and $\theta$. A function $\tilde{m}(u,r,\theta)$ defined by
\begin{equation}
V=e^{2\beta}\left(  r-\frac{2\tilde{m}(u,r,\theta)r^{2}}{r^{2}+\tilde{\alpha
}^{2}\cos^{2}{\theta}}\right)  \;, \label{eq_massin}%
\end{equation}
is the generalization, inside the distribution, of the ``mass
aspect'' defined by Bondi and collaborators
\cite{BondiVandenburgMetzner1962} and in the static limit
coincides with the Schwarzschild mass.

\subsection{Energy-momentum tensor}

It is assumed that, for a local observer co-moving with a fluid having a
velocity $\vec{\omega}=\left(  \omega_{x},0,\omega_{z}\right)  $, the
space-time contains:

\begin{itemize}
\item  an isotropic (pascalian) fluid represented by $\hat{T}_{\mu}^{M\ \nu
}\nolinebreak =\nolinebreak {diag}\,(\rho,-P,-P,-P)$. Where $\rho$
is the energy density and $P=P_{r}$ the radial pressure. Although
the perfect pascalian fluid assumption (i.e. $P_{r}=P_{\perp}$) is
supported by solid observational and theoretical grounds, an
increasing amount of theoretical evidence strongly suggests that,
for certain density ranges, a variety of very interesting physical
phenomena may take place giving rise to local anisotropy (see
\cite{HerreraSantos1997} and references therein).

\item  a radiation field of specific intensity $\mathbf{I}(r,t;\vec{n},\nu)$
given through
\begin{equation}
\mathrm{d}\mathcal{E}=\mathbf{I}(r,t;{\vec{n}},{\nu})\mathrm{d}S\ \cos
\varphi\ \mathrm{d}\Theta\ \mathrm{d}\upsilon\ \mathrm{d}t,
\end{equation}
with $\varphi$ the angle between ${\vec{n}}$ and the normal to $\mathrm{d}S$
and where $\mathrm{d}\mathcal{E}$ is defined as the energy transported by a
radiation of frequencies $\left(  \nu,\nu+\mathrm{d}\upsilon\right)  $ in time
$\mathrm{d}t,$ crossing a surface element $\mathrm{d}S,$ through the solid
angle around ${\vec{n},}$ i.e. $\mathrm{d}\Theta\nolinebreak \equiv
\nolinebreak \sin\theta\mathrm{d}\theta\mathrm{d}\psi\nolinebreak
\equiv\nolinebreak -\mathrm{d}\mu\mathrm{d}\psi$. \newline As in classical
radiative transfer theory, for a planar geometry the moments of $\mathbf{I}%
(r,t;\vec{n},\nu)$ can be written as
\cite{Lindquist1966,MihalasMihalas1984,RezzollaMiller1994}
\begin{equation}
\rho_{R}=\frac{1}{2}\int_{0}^{\infty}d\nu\hspace{0.25cm}\int_{1}^{-1}%
d\mu\hspace{0.25cm}\mathbf{I}(r,t;\vec{n},\nu),\qquad\mathcal{F}=\frac{1}%
{2}\int_{0}^{\infty}d\nu\hspace{0.25cm}\int_{1}^{-1}d\mu\hspace{0.25cm}%
\mu\ \mathbf{I}(r,t;\vec{n},\nu)\label{eq_0mom}%
\end{equation}
and
\begin{equation}
\mathcal{P}=\frac{1}{2}\int_{0}^{\infty}d\nu\hspace{0.25cm}\int_{1}^{-1}%
d\mu\hspace{0.25cm}\mu^{2}\mathbf{I}(r,t;\vec{n},\nu)\hspace{0.25cm}%
.\label{eq_2mom}%
\end{equation}
Physically, $\rho_{R}$ , $\mathcal{F}$ and $\mathcal{P}$, represent the
radiation contribution to the: energy density, energy flux density and radial
pressure, respectively.
\end{itemize}

From the above assumptions the energy momentum tensor can be written as
$\hat{T}_{\mu\nu}\nolinebreak =\nolinebreak \hat{T}_{\mu\nu}^{M}%
\nolinebreak +\nolinebreak \hat{T}_{\mu\nu}^{R}$ where the
material part is $\hat{T}_{\mu\nu}^{M\ }$ and the corresponding
term for the radiation field, $\hat{T}_{\mu\nu}^{R}$, can be
written as \cite{Lindquist1966,MihalasMihalas1984}:
\begin{equation}
\hat{T}_{\mu\nu}^{R}=\left(
\begin{array}
[c]{cccc}%
\rho_{R} & -\mathcal{F} & 0 & 0\\
-\mathcal{F} & \mathcal{P} & 0 & 0\\
0 & 0 & \frac{1}{2}(\rho_{R}-\mathcal{P)} & 0\\
0 & 0 & 0 & \frac{1}{2}(\rho_{R}-\mathcal{P})
\end{array}
\right)  \;. \label{eq_tradiat}%
\end{equation}
Notice the induced anisotropy in the $\hat{T}_{\mu\nu}^{R}$ due to the
radiation field.

Then, the energy-momentum tensor in the local co-moving frame takes the
following form:
\begin{align}
\hat{T}_{{\mu}{\nu}} &  =\left[  \left(  \rho+{\rho}_{R}\right)  +P+\frac
{1}{2}\left(  \rho_{R}-\mathcal{P}\right)  \right]  \hat{U}_{\mu}\hat{U}_{\nu} \nonumber -\left(  P+\frac{1}{2}\left(  \rho_{R}-\mathcal{P}\right)  \right)  \eta_{\mu\nu} \nonumber   + \frac{1}{2}(3\mathcal{P}-\rho_{R})\hat{\chi}_{\mu} \hat{\chi}_{\nu} +\hat{F}_{\mu}\hat{U}_{\nu}+\hat{F}_{\nu}\hat{U}_{\mu
}\;; \nonumber
\end{align}
where $\eta_{\mu\nu}=diag\left(  1,-1,-1,-1\right)  ,\hat{U}_{\mu}=(1,0,0,0)$,
$\hat{\chi}_{\mu}=(0,1,0,0)$ and $\hat{F}_{\mu}=(0,-\mathcal{F},0,0)$.

Now following \cite{HerreraEtal1994}, in order to find the energy
momentum tensor as seen by this observer co-moving with the fluid,
we should perform an infinitesimal rotation around the symmetry
axis, i.e.
\begin{equation}
\bar{T}_{\mu\nu}=\left(
\begin{array}
[c]{cccc}%
\rho+\rho_{R} & -\mathcal{F} & 0 & \frac{1}{2}\mathcal{D}\left(  3\rho
_{R}-\mathcal{P}\right) \\
-\mathcal{F} & P+\mathcal{P} & 0 & -\mathcal{DF}\\
0 & 0 & P+\frac{1}{2}\left(  \rho_{R}-\mathcal{P}\right)  & 0\\
\frac{1}{2}\mathcal{D}\left(  3\rho_{R}-\mathcal{P}\right)  & -\mathcal{DF} &
0 & P+\frac{1}{2}\left(  \rho_{R}-\mathcal{P}\right)
\end{array}
\right)  \;, \label{TEMExplic1}%
\end{equation}
where $\mathcal{D}\left(  u,r,\theta\right)  $ is associated with the local
``dragging of inertial frames'' effect, which in the slow rotation limit
$\mathcal{D}$ will also be taken up to first order.

Once minkowskian co-moving energy momentum tensor is built in terms of
physical observables on a local frame ($\rho$, $P$, $\rho_{R}$ ,
$\mathcal{F},$ $\mathcal{P},$and $\mathcal{D}$), it can be transformed from
the local minkowskian co-moving coordinates $(t,x,y,z)$ to the curvilinear not
co-moving Bondi coordinates $(u,r,\theta,\phi)$ as
\begin{equation}
T_{\alpha\beta}=\frac{\partial\hat{x}^{\gamma}}{\partial x^{\alpha}}%
\;\frac{\partial\hat{x}^{\lambda}}{\partial x^{\beta}}\;L_{\gamma}^{\mu}%
(\vec{\omega})\;L_{\lambda}^{\nu}(\vec{\omega})\;\bar{T}_{\mu\nu}\;;
\label{eq_transftmunu}%
\end{equation}
where $L_{\lambda}^{\nu}(\vec{\omega})$ is a Lorentz boost, written as
\begin{equation}
L_{\gamma}^{\mu}(\vec{\omega})=\left(
\begin{array}
[c]{cccc}%
\gamma & -\gamma\omega_{x} & 0 & -\gamma\omega_{z}\\
-\gamma\omega_{x} & 1+\frac{\omega_{x}^{2}\left(  \gamma-1\right)  }%
{\omega^{2}} & 0 & \frac{\omega_{x}\omega_{z}\left(  \gamma-1\right)  }%
{\omega^{2}}\\
0 & 0 & 1 & 0\\
-\gamma\omega_{z} & \frac{\omega_{x}\omega_{z}\left(  \gamma-1\right)
}{\omega^{2}} & 0 & 1+\frac{\omega_{z}^{2}\left(  \gamma-1\right)  }%
{\omega^{2}}%
\end{array}
\right), \quad \text{with } \gamma=\frac{1}{\sqrt{1-\omega^{2}}}
\quad \text{and} \quad \omega^{2}=\omega_{x}^{2}+\omega_{z}^{2}.  \label{Lorentz}%
\end{equation}

Observe that $\partial\hat{x}^{\gamma}/\partial x^{\alpha}\;$are coordinate transformations connecting $(t,x,y,z)$ with $(u,r,\theta,\phi)$ which can be identify from equations (\ref{minKdt}) through (\ref{minKdxdydz}).

In radiation coordinates the radial and orbital velocities of matter are given by
\begin{equation}
\frac{\mathrm{d}r}{\mathrm{d}u}=\frac{V}{r}:\frac{\omega_{x}}{1-\omega_{x}%
}\hspace{0.5cm}\text{and}\hspace{0.5cm}\frac{\mathrm{d}\phi}{\mathrm{d}%
u}=\frac{\omega_{z}}{1-\omega_{x}}\frac{1}{r\sin\left(  \theta\right)
}e^{\beta}\sqrt{\frac{V}{r}}, \label{eq_veloc}%
\end{equation}
respectively.

Now, using the metric (\ref{minterna}), the energy momentum tensor (\ref{TEMExplic1}), the transformation (\ref{eq_transftmunu}) and considering the slow rotation limit (i.e. first order in the orbital velocity $\omega_{z},$ Kerr parameter $\tilde{\alpha}$ and the dragging function $\mathcal{D}$), we can write the Einstein (see Appendix \ref{EinsteinEquations}). As in reference \cite{HerreraEtal1994} only six of the eight physical variables
($\omega_{x}, \omega_{z}, \rho, P, \rho_{R},\mathcal{F},\mathcal{P}$ and $\mathcal{D}$), can be algebraically obtained, in terms of the metric functions $\beta(  u,r,\theta)  $ and $\tilde{m}( u,r,\theta)  $ and their derivatives, from field equations (\ref{Tuu}) through (\ref{Tup}) . Therefore, more information (equations) has to be provided to this system in order to solve the physical variables. The idea will be to supply relations among the radiation physical variables $\rho_{R},\mathcal{F},$ and $\mathcal{P.}$ Next section will be devoted to describe these
essential relations.

\section{Closures relations and the limits for the radiation field}
\label{eddingtonfactor}
In order to deal with more realistic scenarios, the microphysical framework of the interrelation between matter and radiation have to be considered. The relativistic Boltzmann Transport Equation must be coupled to the hydrodynamic equations in order to obtain the evolution of the system as well as the spectrum and angular distribution of the radiation field \cite{Lindquist1966} Neglecting effects as polarization, dispersion and coherence, a covariant special relativistic equation of\ radiation transport has been proposed as \cite{AndersonSpiegel1972,AliRomano1994}:
\begin{equation}
\left(  u^{\mu}+l^{\mu}\right)  \left\{  \mathbf{\nabla}_{\mu}\ \mathbf{I}%
+4\mathbf{I}\ l^{\sigma}\ \mathbf{\nabla}_{\mu}\ u_{\sigma}+l^{\rho
}\ l^{\sigma}\ \frac{\partial\ \mathbf{I}}{\partial\ l^{\rho}}\ \mathbf{\nabla
}_{\mu}\ u_{\sigma}+u^{\rho}\ l^{\sigma}\frac{\partial\ \mathbf{I}}%
{\partial\ l^{\rho}}\ \mathbf{\nabla}_{\mu}\ u_{\sigma}-\frac{\partial
\ \mathbf{I}}{\partial\ l^{\rho}}\ \mathbf{\nabla}_{\mu}\ u^{\rho}\right\}
=\rho(\epsilon_{0}-\kappa I)\ ,\label{ealirom}%
\end{equation}
where $l_{\mu}l^{\mu}=1$ with $l_{\mu}u^{\mu}=0;$ the four
velocity of the fluid is $u^{\mu}$; $\rho$ is proper density of
the medium; the quantities $\epsilon_{0}$ and $\kappa$ are the
emissivity and the absorption coefficient, respectively. This
transfer equation has several important difficulties. The most
important are: the lack of information about the coupling between
radiation and ultradense matter and its mathematical complexity,
although some understanding is emerging
recently\cite{EfimovEtal1997,WehrseBaschek1999}.

One of the possible strategies to circumvent the difficulty of
solving the radiation transfer equation is to consider\ one of the
two physical reasonable limits for the radiation field\textit{ }
which describe a significant variety of astrophysical
scenarios\cite{MihalasMihalas1984}. The\textit{\ free streaming
out }limit assumes that radiation (neutrinos and/or photons) mean
free path is of the order of the dimension of the sphere. This was
the case considered in \cite{HerreraEtal1994} and it can be
expressed as
\begin{equation}
\rho_{R}=\mathcal{F}=\mathcal{P}=\hat{\epsilon}\;. \label{free_streaming}%
\end{equation}
The other limit for the radiation field is the\textit{\ diffusion limit
approximation,} where radiation is considered to flow with a mean free path
much smaller than the characteristic length of the system. Within this limit,
radiation is locally isotropic and we have
\begin{equation}
\rho_{R}=3\mathcal{P}\mathrm{\hspace{0.5cm}}\text{and}\hspace{0.5cm}%
\mathcal{F}={\hat{q}}\;. \label{diffusio_appox}%
\end{equation}

In order to simulate more realistically the matter and radiation interaction,
it seems more reasonable to have a parameter which varies between the above
mentioned limits. This is the idea of the flux and the variable Eddington
factor an they can be summarized as follows. From equations (\ref{eq_0mom})
through (\ref{eq_2mom}) it is convenient to define the following normalized quantities%

\begin{equation}
\varphi(\vec{r},t,\Omega)=\frac{I(\vec{r},t,\Omega)}{\rho_{R}},\mathrm{\quad
}\mathbf{\vec{f}}=\int_{4\pi}\varphi(\vec{r},t,\Omega)\vec{n}d\Omega
\mathrm{\quad}\text{and}\mathrm{\quad}\mathbb{K=}\int_{4\pi}\varphi(\vec
{r},t,\Omega)\vec{n}\otimes\vec{n}d\Omega, \label{cn}%
\end{equation}
The Eddington factor is, defined as the eigenvalue of the Pressure Tensor corresponding to the eigenvector $\vec{n}$ (unitary vector in the direction of the energy flux ), i.e. $K_{j}^{i}n^{j}=\chi n^{i}$ \cite{AnilePennisiSammartino1991}. Thus,
\begin{equation}
\text{\ }\mathbf{\vec{f}}\Rightarrow f^{i}=fn^{i}\mathrm{\quad}\text{and}%
\mathrm{\quad}\mathbb{K\Rightarrow\ }K^{ij}=\frac{1}{2}\left\{  (1-\chi
)\delta^{ij}+(3\chi-1)n^{i}n^{j}\right\}  \ . \label{anile1}%
\end{equation}
In the one-dimensional case the above equations lead to%
\begin{equation}
f=\frac{\mathcal{F}}{\rho_{R}\text{ }}\mathrm{\quad}\text{and}\mathrm{\quad
}\mathcal{\chi}=\frac{\mathcal{P}}{\rho_{R}}\ . \label{anile2}%
\end{equation}
Which are called the flux and the variable Eddington factor, respectively.

In order to ``close'' this problem and to algebraically obtain six of the
above mentioned physical variables, namely $\omega_{x},$ $\omega_{z},$ $\rho,$
$P,$ $\mathcal{F},$ and $\mathcal{D}$, from field equations (\ref{Tuu})
through (\ref{Tup}) and the radiation parameter (\ref{anile2}) (or in general
(\ref{cn})) we need to state a relation between $f$, and $\chi$. It is easy to
perceive that such a relation could exist. In fact, it is noticeable that in
the corresponding limits for the radiation field, i.e. \textit{diffusion limit
approximation} and \textit{free streaming out} we have%
\begin{equation}
\left.
\begin{array}
[c]{c}%
\mathcal{P}=\frac{1}{3}\rho_{R}\Rightarrow f\longrightarrow0\mathrm{\quad
}\text{and}\mathrm{\quad}\chi=\frac{1}{3}\\
\\
\mathcal{F=}\ \mathcal{P=}\rho_{R}\Rightarrow f=1\mathrm{\quad}\text{and}%
\mathrm{\quad}\chi=1
\end{array}
\right\}  \Rightarrow0\leq f\leq1\mathrm{\quad}\text{and}\mathrm{\quad}%
\frac{1}{3}\text{\ }\leq\chi(f)\leq1 \label{efeji}%
\end{equation}
Causality requirement implies the following supplementary
conditions on $f$ and $\chi,$ in order to define a physically
plausible region in the $\left\{
f,\chi,{\mathrm{d}\chi}/{\mathrm{d}f}\right\}  $ space
\cite{PonsIbanezMiralles2000}%
\begin{equation}
\left\|  f\right\|  \leq1,\mathrm{\quad}f^{2}\leq\chi\leq1\mathrm{\quad
}\text{and}\mathrm{\quad}-\frac{1-\chi}{1+f}\leq\frac{\mathrm{d}\chi
}{\mathrm{d}f}\leq\frac{1-\chi}{1-f} \label{condfdxi}%
\end{equation}%

\begin{table}[h] \centering
\begin{tabular}
[c]{|c|c|c|c|}\hline\hline 
\textbf{Closure} & $\chi(f)  $ & $\left.  \frac{\mathrm{d}\chi}{\mathrm{d}f}\right|  _{f=1}$ & $\left.  \frac{\mathrm{d}\chi}{\mathrm{d}f}\right|  _{f=0}$ \\ \hline\hline
\multicolumn{1}{|l|}{\textit{Lorentz-Eddington }(LE)} & $\frac{5}{3}-\frac {2}{3}\sqrt{4-3f^{2}}$ & $2$ & $0$\\\hline
\multicolumn{1}{|l|}{\textit{Bowers-Wilson}} & $\frac{1}{3}\left(
1-f+3f^{2}\right)  $ & $\frac{5}{3}$ & $-\frac{1}{3}$\\\hline
\multicolumn{1}{|l|}{\textit{Janka (Monte Carlo) }(MC)} &
$\frac{1}{3}\left(
1+\frac{1}{2}f^{1.31}+\frac{3}{2}f^{4.13}\right)  $ & $2.28$ &
$0$\\\hline \multicolumn{1}{|l|}{\textit{Maximum Packing} (MP)} &
$\frac{1}{3}\left( 1-2f+4f^{2}\right)  $ & $2$ &
$-\frac{2}{3}$\\\hline \multicolumn{1}{|l|}{\textit{Minerbo} (Mi)}
& $\chi\left(  f\right)
=1-2\frac{f}{\kappa}\quad$ where $\quad f=\coth\kappa-\frac{1}{\kappa}$ & $2$%
& $0$\\\hline \multicolumn{1}{|l|}{\textit{Levermore-Pomraning}} &
$\chi\left(  f\right) =f\coth\beta\quad$where$\quad$ $f=
\coth\beta- \frac{1}{\beta}$ & $1$ & $0$\\\hline
\end{tabular}
\caption{Closure Relations and some of their physical acceptability conditions}
\label{tabla1}
\end{table}

There are several closure relations reported in the literature (see two recent comprehensive discussions on this subject in \cite{PonsIbanezMiralles2000} and \cite{SmitVandenHornBludman2000} and references therein). Few of them are simply \textit{ad hoc} relations that smoothly interpolate the radiation field between the diffusive and free-streaming regimes. Others, are derived from a maximum entropy principle or from a given, or assumed, angular dependence of the radiative distribution functions. Even one of them has been motivated from direct transport calculations. Six of the most frequent found closure relations are listed in Table \ref{tabla1}. In this list the first four could be considered as ``analytical'' closure relations, while the last two are referred as numerical, because for a given flux factor $f,$ the nonlinear equation $f=\coth\beta-({1}/{\beta})$ has to be numerically solved in order to obtain the variable Eddington factor $\chi$.

We are going to explore some of the effects of dissipation on the evolution of slowing rotating radiating matter configuration in General Relativity. We shall evaluate how independent are these effects from an explicit closure relation and/or a specific EoS chosen. Particularly, some results concerning the influence of the junction conditions on the eccentricity and the radiation scheme evaluated at the surface, will be presented in the next section. The strategy we follow to close the system of Einstein field equations with a radiation field, contrasts with the standard iterative method for solving the moment equations (\ref{eq_0mom}), and (\ref{eq_2mom}), starting from an estimated Eddington factor (see \cite{MihalasMihalas1984} and \cite{RamppJanka2002} and references therein).

Thus, by using equations (\ref{anile2}), the Einstein field equations (\ref{Tuu})-(\ref{Tup}) can be re-written in terms of the flux and the Eddington factors (see Appendix \ref{EinsteinFluxFactor}) . Again, in principle, for all cases listed in Table \ref{tabla1} (including numerical closures relations) it is possible to obtain the remaining six physical variables, $\omega_{x}, \omega_{z}, \rho, P, \mathcal{F},$ and $\mathcal{D}$, from the system (\ref{TEdduu}) through (\ref{TEddup}), in terms of the Kerr parameter, $\overset{\thicksim}{\alpha},$ the flux factor, $f$ , the metric functions $\beta( u,r,\theta), \tilde{m}( u,r,\theta)  $ and their derivatives.

\section{Eddington factor, flux factor and junction conditions}

\label{JunctionConditions}In this section, following \cite{HerreraEtal1994} we should match the interior fluid
spheroid to the exterior Kerr-Vaidya solution (equation (\ref{mexterna})). Therefore, the continuity of the \textit{first}
and the \textit{second fundamental} ($g_{ij}$ and $K_{ij}$) forms across the matching surface are needed. These requirements are equivalent to demand the continuity of the \textit{tetrad components} and \textit{spin coefficient} of the metrics (\ref{mexterna}) and (\ref{minterna}) across the boundary surface  $r=a\left(  u\right)  $ \cite{HerreraJimenez1983}.

\subsection{Junction conditions for a slowly rotating configuration}

The Newman-Penrose null tetrad components for the metrics (\ref{mexterna}) and (\ref{minterna}) (see reference
\cite{HerreraEtal1994} to find the expressions of the spin coefficients for these two metrics) are for the exterior metric
\begin{equation}
l^{\mu}  =\delta_{r}^{\mu};\quad n^{\mu}=\delta_{u}^{\mu}-\frac{1}{2}\left[  1-\frac{2mr}{r^{2}+\alpha^{2}\cos^{2}\left(  \theta\right)}\right]  \delta_{r}^{\mu}; \quad 
m^{\mu}  =\frac{ \left[  i\alpha\sin\theta\left(  \delta_{u}^{\mu}-\delta_{r}^{\mu}\right)  +\delta_{\theta}^{\mu}+i\csc\theta\delta_{\phi}^{\mu}\right]}{\sqrt{2}\left(  r+i\alpha\cos \theta \right)  };
\end{equation}
and 
\begin{equation}
l^{\mu}  =e^{-2\beta}\delta_{r}^{\mu};\quad 
n^{\mu}=\delta_{u}^{\mu}-\frac{1}{2}e^{-2\beta}\left[  1-\frac{2\tilde{m}r}{r^{2} + \widetilde{\alpha}^{2}\cos^{2} \theta  }\right]  \delta_{r}^{\mu} \quad 
m^{\mu}   =\frac{\left[  i\widetilde{\alpha}\sin\theta\left(  \delta_{u}^{\mu}-\delta_{r}^{\mu}\right)  +\delta_{\theta}^{\mu}+i\csc\theta \delta_{\phi}^{\mu}\right]}{\sqrt{2}\left(  r+i\widetilde{\alpha}\cos\theta\right)  }  .
\end{equation}
for the interior metric (\ref{minterna})

The continuity of the tetrad components across the boundary surface $r=a(u)$ implies
\begin{equation}
\beta_{a}=0;\qquad\tilde{m}_{a}=m\qquad\text{and}\qquad\widetilde{\alpha}_{a}=\alpha\ , \label{beta-eme-alpha}%
\end{equation}
and the continuity of the spin coefficients $\tau,\gamma,$ and $\nu$ lead to
\begin{equation}
\beta_{1a}\left(  1-\frac{2m}{a}\right)  -\beta_{0a}=\frac{\tilde{m}_{1a}}{2a}; \quad \beta_{2a}=\tilde{m}_{2a}=0 \quad \text{and } \alpha \left(  \beta_{1a}-\beta_{0a}\right)  =\alpha\left(  m_{0a}-\tilde{m}_{0a}+\tilde{m}_{1a}\right)  =0, \label{spinTau-Gamma}%
\end{equation}%
which means that $\left(  \beta_{1a}-\beta_{0a}\right)  $ and $\left(m_{0a}-\tilde{m}_{0a}+\tilde{m}_{1a}\right)  $ are of order $\alpha.$

Now, evaluating the field equations (\ref{Tuu}) - (\ref{Tup}) (or equivalently (\ref{TEdduu}) through (\ref{TEddup}), ) at $r=a(u)$ and considering the above results (\ref{beta-eme-alpha}), we obtain that, on these coordinates and up to the first order in $\alpha,$ the metric coefficients $\beta$ and $\tilde{m}$ are \textit{independent of the angular variable} and consequently the physical variables: $\omega_{x},$ $\rho,$ $P,$ and $\mathcal{F}$ are also $\theta-$independent (see \cite{HerreraEtal1994} for details).

Next, expanding $\beta$ near the surface, $\beta_{0a}+\overset{.}{a}\beta_{1a}=0$, in the two first equations (\ref{spinTau-Gamma}) and using that $\beta$ is continuous and vanishes at the outside the matter configuration we obtain that
\begin{equation}
\beta_{1a}\left(  1-\frac{2\ \tilde{m}}{a}\right)  -\beta_{0a}=\frac{\tilde
{m}_{1a}}{2a}\Longrightarrow\dot{a}=\left(  1-\frac{2\tilde{m}_{a}}{a}\right)
\left[  \frac{\left(  \rho_{a}+\rho_{Ra}-\mathcal{F}_{a}\right)  \omega
_{xa}-\left(  P_{a}+\mathcal{P}_{a}-\mathcal{F}_{a}\right)  }{\left(  \rho
_{a}+\rho_{Ra}+P_{a}+\mathcal{P}_{a}-2\mathcal{F}_{a}\right)  \left(
1-\omega_{xa}\right)  }\right]  \ , \label{dadu1}%
\end{equation}
where $\dot{a}=\mathrm{d}r/\mathrm{d}u.$ On the other hand, from equation
(\ref{eq_veloc}) it follows that
\begin{equation}
\dot{a}=\left(  1-\frac{2\tilde{m}_{a}}{a}\right)  \frac{\omega_{xa}}%
{1-\omega_{xa}}. \label{dadu2}%
\end{equation}
Equating (\ref{dadu1}) and (\ref{dadu2}), it is obtained that the emerging energy flux density compensates the total pressure (hydrodynamic and radiation) inside the configuration \cite{AguirreHernandezNunez1994}, i.e.
\begin{equation}
\mathcal{F}_{a}=P_{a}+\mathcal{P}_{a}\ . \label{pressurea}%
\end{equation}

Now, expanding $\beta$ and $\tilde{m}$ near the surface in the third equation (\ref{spinTau-Gamma}), we conclude that
\begin{equation}
\beta_{1a}\left(  1+\dot{a}\right)  \approx\alpha\Longrightarrow\beta_{1a}\left(  1+\dot{a}\right)  =v(u)\alpha 
\quad \text{and} \quad
\tilde{m}_{1a}\left(  1+\dot{a}\right)  \approx\alpha\Longrightarrow\tilde{m}_{1a}\left(  1+\dot{a}\right)  =q(u)\alpha\ . \label{definiefe-ge}%
\end{equation}
Where $v(u)$ and $q(u)$ are arbitrary functions of the time-like coordinate
$u$ with $\left|  v(u)\right|  \lessapprox1$ and $\left|  q(u)\right|
\lessapprox1$ in order to keep valid the approximation. Additionally, by using
field equations (\ref{Tur}) and (\ref{Trr}), we get that%
\begin{equation}%
\begin{array}[c]{ccc}%
\beta_{1a}\left(  1+\dot{a}\right)  \approx\alpha & \Longrightarrow &
\dfrac{2\pi a\left(  1-\omega_{xa}\dfrac{2\ \tilde{m}_{a}}{a}\right)
}{\left(  1-\dfrac{2\ \tilde{m}_{a}}{a}\right)  \left(  1+\omega_{xa}\right)
}\left[  \rho_{a}+\rho_{Ra}-\mathcal{F}_{a}\right]  \approx\alpha\ \\
\text{and} &  & \\
\tilde{m}_{1a}\left(  1+\dot{a}\right)  \approx\alpha & \Longrightarrow &
\dfrac{a^{2}4\pi}{1+\omega_{xa}^{2}}\left(  1-\omega_{xa}\dfrac{2\ \tilde
{m}_{a}}{a}\right)  \left[  \rho_{a}+\rho_{Ra}-\mathcal{F}_{a}\right]
\approx\alpha\ ,
\end{array}
\label{restricordvarfis}%
\end{equation}
which impose restrictions on the physical variable evaluated at the surface of
the distribution.

Also, it follows from the junction conditions (\ref{spinTau-Gamma}) that
\begin{equation}
2a\beta_{1a}\left(  1+\dot{a}-\frac{2\tilde{m}_{a}}{a}\right)  =\tilde{m}%
_{1a}\quad\Longleftrightarrow\quad2a\beta_{1a}\left(  1-\frac{2\tilde{m}_{a}%
}{a}\right)  =\tilde{m}_{1a}\left(  1-\omega_{xa}\right)  . \label{segforma2}%
\end{equation}
Thus, we obtain an expression relating $v(u)$ and $q(u),$ namely
\begin{equation}
2a\left(  1+\dot{a}-\frac{2\tilde{m}_{a}}{a}\right)  v(u)=q(u)\qquad
\Longleftrightarrow\qquad\frac{2a}{\left(  1-\omega_{xa}\right)  }\left(
1-\frac{2\tilde{m}_{a}}{a}\right)  v(u)=q(u). \label{relaefege}%
\end{equation}
However, it can be checked by simple inspection that, because neither the
field equations nor the junction conditions impose further limitations on
these functions of $u$, one of them remains completely arbitrary for each model.

More over, at least for some models we have
\begin{equation}
\frac{2a}{\left(  1-\omega_{xa}\right)  }\left(  1-\frac{2\tilde{m}_{a}}%
{a}\right)  \sim1,\label{orderparameter}%
\end{equation}
which becomes useful when selecting the order of magnitude of the initial
parameters ($\tilde{m}_{a},a$ and $\omega_{xa}$) for the modelling of slowly
rotating collapsing configurations (i.e. first order in the orbital velocity
$\omega_{z},$ Kerr parameter $\tilde{\alpha}$ and the dragging function
$\mathcal{D}$), worked out in section \ref{modelingscenarios}.

The next section will be devoted to explore consequences of the radiation
field, $T_{\mu\nu}^{R},$ in collapsing configurations in the slow rotation approximation.

\subsection{The limits for the flux factor and the eccentricity}

\begin{figure}
\begin{center}
\includegraphics[width=5in]{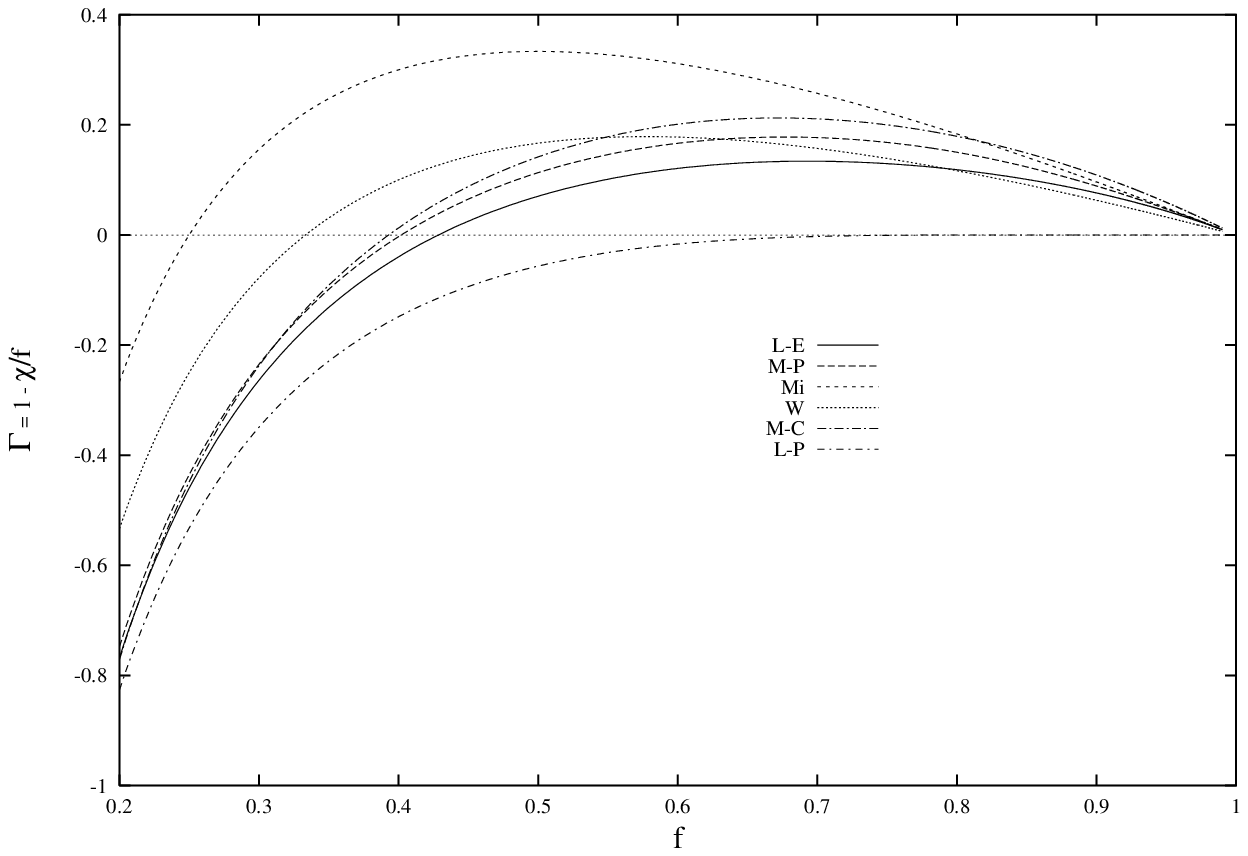}%
\caption{The $\left(  1-\frac{\chi_{a}}{f_{a}}\right)  $ vs $f$
for the different closure relations. \textbf{L-E}$\rightarrow$ \textit{Lorentz-Eddington};
\textbf{M-P}$\rightarrow$\textit{Maximum Packing;} \textbf{Mi}$\rightarrow$\textit{Minerbo; }\textbf{W}$\rightarrow
$\textit{Bowers-Wilson; }\textbf{M-C}$\rightarrow$\textit{Janka
(Monte Carlo);
}\textbf{L-P}$\rightarrow$\textit{Levermore-Pomraning.}}%
\label{f1}%
\end{center}
\end{figure}

Now, from equations (\ref{anile2}), and (\ref{pressurea}), we have
\begin{equation}
P_{a}=\mathcal{F}_{a}\left(  1-\frac{\mathcal{P}_{a}}{\mathcal{F}_{a}}\right)
\ \qquad\Longrightarrow P_{a}=\mathcal{F}_{a}\left(  1-\frac{\chi_{a}}{f_{a}%
}\right)  . \label{preshydro}%
\end{equation}
If we assume that the hydrodynamic pressure and the outgoing energy flux have to be positive then we have that $\left( 1- ({\chi_{a}}/{f_{a}}) \right) \geq0$. Figure \ref{f1} displays this factor for the different closure relations in Table \ref{tabla1}. It is clear from this figure that the boundary conditions compel the impossibility to attain total diffusion regime\ ( i.e.$\ f=0$)\ at the surface of the configuration. The roots of each curve representing a closure relation define the interval of acceptability for the values of the flux factor $f$ . These intervals are displayed in the second column of Table \ref{tabla2}. As it can be appreciated form Figure \ref{f1}, only
the \textit{Levermore-Pomraning} closure relation does not meet
this requirement. Up to the precision of our numerical
calculation, the acceptable value for $f$ that guarantees the
positiveness of the hydrodynamic pressure is $1$. Thus,
considering the \textit{Levermore-Pomraning} closure relation,
junction conditions, up to the first order in Kerr rotation
parameter, only allows free streaming out at the surface for this
slowly rotating matter distribution. On the other hand,
\textit{Minerbo} closure relation seems to admits transport
mechanism closer to the diffusion limit.

\begin{table}[tbp] \centering
\begin{tabular}
[c]{|c|c|c|}\hline\hline \multicolumn{1}{||c|}{\textbf{Closure}} &
\multicolumn{1}{||c|}{$f_{r=a}$} &
\multicolumn{1}{||c||}{$e$}\\\hline\hline
\multicolumn{1}{|l|}{\textit{Lorentz-Eddington}} &
$\frac{3}{7}\leq\left. f_{LE}\right|  _{r=a}\leq1$ & $\Lambda\leq
e_{LE}\leq\Lambda\left(
1+\dfrac{4}{3}\dfrac{\mathcal{F}_{a}}{\rho_{a}}\right)  $\\\hline
\multicolumn{1}{|l|}{\textit{Bowers-Wilson}} &
$\frac{1}{3}\leq\left. f_{BW}\right|  _{r=a}\leq1$ & $\Lambda\leq
e_{BW}\leq\Lambda\left(
1+2\dfrac{\mathcal{F}_{a}}{\rho_{a}}\right)  $\\\hline
\multicolumn{1}{|l|}{\textit{Janka (Monte Carlo)}} &
$0.39\leq\left. f_{MC}\right|  _{r=a}\leq1$ & $\Lambda\leq
e_{MC}\leq\Lambda\left(
1+1.545\dfrac{\mathcal{F}_{a}}{\rho_{a}}\right)  $\\\hline
\multicolumn{1}{|l|}{\textit{Maximum Packing}} &
$\frac{1}{4}\leq\left. f_{MP}\right|  _{r=a}\leq1$ & $\Lambda\leq
e_{MP}\leq\Lambda\left(
1+3\dfrac{\mathcal{F}_{a}}{\rho_{a}}\right)  $\\\hline
\multicolumn{1}{|l|}{\textit{Minerbo}} & $0.40\leq\left.
f_{M}\right| _{r=a}\leq1$ & $\Lambda\leq e_{M}\leq\Lambda\left(
1+1.488\dfrac {\mathcal{F}_{a}}{\rho_{a}}\right)  $\\\hline
\multicolumn{1}{|l|}{\textit{Levermore-Pomraning}} & $\left.
f_{LP}\right| _{r=a}=1$ & $e_{LP}=\Lambda$\\\hline
\end{tabular}
\caption{Limits for the flux factors and the eccentricity for the different closure relations}
\label{tabla2}
\end{table}

Because all these results emerge from the junction conditions that couple the
internal and the external solutions, they are valid not only for axisymmetric
configurations but also for spherical ones. It is also independent of the EoS
and is present for all the closure relations we have listed in the Table
\ref{tabla1}.

Finally, expanding (\ref{eccentricity}) for $\tilde{\alpha}\ll1$ ,
we get $\tilde{\alpha}$\cite{HerreraEtal1998}
\begin{equation}
e=\frac{1}{r_{s}}\tilde{\alpha}-\frac{1}{2}\frac{1}{r_{s}^{3}}\tilde{\alpha
}^{3}+\cdots\ , \label{expeccen}%
\end{equation}
as expected, up to first order, the eccentricity is proportional to
$\tilde{\alpha}.$ Now, using the field equation (\ref{Tur}) evaluated at the
surface $r=a\left(  u\right)  $, (\ref{definiefe-ge}) and (\ref{relaefege}) we
are lead to%
\begin{equation}
\tilde{\alpha}_{a}=\alpha=\frac{2\pi a\left(  \rho_{a}+\mathcal{F}_{a}%
\dfrac{\left(  1-f_{a}\right)  }{f_{a}}\right)  \left(  1-\omega_{xa}%
\dfrac{2\tilde{m}_{a}}{a}\right)  }{\left(  1+\omega_{xa}\right)  \left(
1-\dfrac{2\tilde{m}_{a}}{a}\right)  v(u)}, \label{alpha}%
\end{equation}
and the surface eccentricity can be re-written as
\begin{equation}
e_{a}=\frac{2\pi\left[  \rho_{a}+\mathcal{F}_{a}\dfrac{\left(  1-f_{a}\right)
}{f_{a}}\right]  \left(  1-\omega_{xa}\dfrac{2\tilde{m}_{a}}{a}\right)
}{\left(  1+\omega_{xa}\right)  \left(  1-\dfrac{2\tilde{m}_{a}}{a}\right)
v(u)}. \label{eccentgeneral}%
\end{equation}
Notice that, $v(u)$ remains completely arbitrary and its choice completes the
characterization of the model.

Because of the range of acceptability for the flux factor we also obtain a
range for the eccentricity, i.e.%
\begin{equation}
f_{\min\chi}\leq f\leq1\quad\Rightarrow
\Lambda\leq e_{a} \leq \Lambda\left(  1+\frac{\mathcal{F}_{a}}{\rho_{a}}\frac{\left(  1-f_{a}\right)
}{f_{a}}\right)  \quad\text{where}\quad\Lambda=\frac{2\pi\rho_{a}\left(
1-\omega_{xa}\dfrac{2\tilde{m}_{a}}{a}\right)  }{\left(  1+\omega_{xa}\right)
\left(  1-\dfrac{2\tilde{m}_{a}}{a}\right)  v(u)}%
\end{equation}
Therefore, it is clear from Table \ref{tabla2} that radiation mechanism
affects the oblateness of the configuration. This is to say, in these
coordinates up to first order in $\tilde{\alpha}_{a}$ and for those models
having $\Lambda>0$, the eccentricity at the surface of a radiating
configuration is greater for models near the diffusion limit approximation
than for those in the free streaming out limit. Again, this result is also
valid for any EoS with $0\eqslantless v(u)\eqslantless1$ and it is present for
all the closure relations in the Table \ref{tabla1}. 

In the next section we shall explore the effect of these closure relations on
the collapse of a radiating, slow-rotating self-gravitating relativistic configuration.

\section{The HJR method and the surface equations}

\label{TheMethod}In order to obtain the evolution of the profiles
of the physical variables, $\omega_{x},$ $\omega_{z},$ $\rho,$
$P,$ $\mathcal{F},$ and $\mathcal{D}$, we use an extension of the
HJR method \cite{HerreraJimenezRuggeri1980} to axially symmetric
slowly rotating case \cite{HerreraEtal1994}.

First, we define two auxiliary variables which, in terms of the Eddington and
the Flux factor, can be written as%
\begin{equation}
\widetilde{\rho}=\frac{\rho+\rho_{R}-\mathcal{F-}\omega_{x}%
(P+\mathcal{P-\mathcal{F})}}{1+\omega_{x}}\equiv\frac{\rho-\omega_{x}%
P+\frac{1}{f}(1-f-\omega_{x}(\chi-f))\mathcal{F}}{1+\omega_{x}}\ ,
\label{densefect}%
\end{equation}
and is called the effective density and, correspondingly, the effective
pressure is%
\begin{equation}
\widetilde{P}=\frac{P+\mathcal{P}-\mathcal{F-}\omega_{x}\left(  \rho+\rho
_{R}-\mathcal{F}\right)  }{1+\omega_{x}}\equiv\frac{P-\omega_{x}\rho+\frac
{1}{f}(\chi-f-\omega_{x}(1-f))\mathcal{F}}{1+\omega_{x}}\ . \label{presefect}%
\end{equation}
With these \textit{effective variables}, the metric elements (equations
(\ref{Tur}) and (\ref{Trr})) can be formally integrated as%
\begin{equation}
\beta\left(  u,r\right)  =\int_{a}^{r}2\pi\overline{r}\frac{\widetilde{\rho
}+\widetilde{P}}{\left(  1-\frac{2\widetilde{m}}{\overline{r}}\right)
}d\overline{r}\qquad\text{and}\qquad\widetilde{m}\left(  u,r\right)  =\int
_{0}^{r}4\pi\ \overline{r}^{2}\ \widetilde{\rho}d\overline{r}\ .
\label{betayeme}%
\end{equation}
Thus, if the $r$ dependence of $\tilde{P}$ and $\tilde{\rho}$ are
known, we can get the metric functions $\tilde{m}$ and $\beta$ up
to some functions of $u$ related to the boundary conditions. This
is one of the key points to transform the Einstein System into a
system of (coupled nonlinear) ordinary differential equations on
the time-like coordinate. Physically, the rationale behind the
assumption on the $r$ dependence of the\textit{\ effective
variables }$\tilde{P}$ and $\tilde{\rho}$, can be grasped in terms
of the characteristic times for different processes involved in a
collapse scenario. If the hydrostatic time scale
$\mathcal{T}_{HYDR}$, which is of the order $\sim1/\sqrt{G\rho}$
(where $G$ is the gravitational constant and $\rho$ denotes the
mean density) is much smaller than the \textit{Kelvin-Helmholtz}
time scale ( $\mathcal{T}_{KH}$ ), then in a first approximation
the inertial terms in the equation of motion can be ignored
\cite{KippenhahnWeigert1990}. Therefore in this first
approximation (quasi-stationary approximation) the $r$ dependence
of $P$ and $\rho$ are the same as in the static solution. Then the
assumption that the\textit{\ effective variables}
(\ref{densefect}) and (\ref{presefect}) have the same $r$
dependence as the physical variables of the static situation,
represents a correction to that approximation, and is
expected to yield good results whenever $\mathcal{T}_{KH}\gg\mathcal{T}%
_{HYDR}$. Fortunately enough,
$\mathcal{T}_{KH}\gg\mathcal{T}_{HYDR}$, for almost all kind of
stellar objects. Recently this rationale becomes intelligible and
finds full justification within the context of a suitable
definition of the post-quasi-static approximation for the
gravitational
collapse\cite{BarretoMartinezRodriguez2002,HerreraEtal2002}.

Those functions of the time-like coordinate $u$ that remain arbitrary can be
obtained from a system of ordinary differential equations (\textit{The System
of Surface Equations, SSE}) emerging from the junction conditions and both the
field equations and some kinematic definitions evaluated at the boundary
surface. The first surface equation is (\ref{dadu2}):
\begin{equation}
\dot{A}=F\left(  \Omega-1\right)  \;. \label{eqSup1}%
\end{equation}
Where we have scaled the radius $a$, the total mass $\tilde{m}_{a}~=~m$ and
the timelike coordinate $u$ by the total initial mass, $m(u~=~0)~=~m\left(
0\right)  $, i.e.
\begin{equation}
A=\frac{a}{m(0)},\hspace{1cm}M=\frac{m}{m(0)},\hspace{1cm}u=\frac{u}{m(0)}
\label{VaRescal}%
\end{equation}
and we have also defined
\begin{equation}
F=1-\frac{2M}{A},\hspace{1cm}\text{and}\hspace{1cm}\Omega=\frac{1}%
{1-\omega_{xa}}\ . \label{FyOmega}%
\end{equation}
Again, the dot over the variable represents the derivative with respect to the
time-like coordinate. The second \textit{Surface Equation} emerges from the
evaluation of equation (\ref{TEdduu}) at $r=a_{+0}$. It takes the form of
\begin{equation}
\dot{M}=-FL\;, \label{eqSup2a}%
\end{equation}
where $L$ representing the total luminosity can be written as
\begin{equation}
L=4\pi A^{2}\mathcal{F}_{a}\left(  2\Omega-1\right)  . \label{eqSup2b}%
\end{equation}
Now, using above equation (\ref{eqSup1}) and definitions (\ref{VaRescal}) and
(\ref{FyOmega}); we can re-state equation (\ref{eqSup2a}) as
\begin{equation}
\frac{\dot{F}}{F}=\frac{2L+(1-F)(\Omega-1)}{A}\;. \label{eqSup2}%
\end{equation}

Finally, after some straightforward manipulations, starting from field
equations (\ref{TEddur}), (\ref{TEddrr}) and (\ref{TEddtt}), it is obtained
\begin{equation}
e^{2\beta}\left(  \frac{\tilde{\rho}+\tilde{P}}{1-\frac{2\tilde{m}}{r}%
}\right)  _{,0}-\frac{\partial\tilde{P}}{\partial r}-\frac{\tilde{\rho}%
+\tilde{P}}{1-\frac{2\tilde{m}}{r}}\left(  4\pi r\tilde{P}+\frac{\tilde{m}%
}{r^{2}}\right)  =\frac{-2}{r}\left(  P+\frac{1}{2}\left(  \rho_{R}%
-\mathcal{P}\right)  -\tilde{P}\right)  , \label{eq_tov}%
\end{equation}
which is the generalization of Tolman-Oppenheimer-Volkov (TOV) equation for any dynamic radiative situation. The third \textit{Surface Equation} can be obtained evaluating (\ref{eq_tov}) at $r~=~a_{+0}$, and it takes the form of:
\begin{equation}
0  =\frac{\dot{\Omega}}{\Omega}+ \frac{\dot{F}}{F}+
\frac{(\tilde{\rho}_{a})_{,0}}{\tilde{\rho}_{a}} +
\frac{F\Omega^{2}\tilde{R}}{\tilde{\rho}_{a}}-\frac{2F \Omega }{A\tilde{\rho}_{a}}\left(  P_{a}+\frac{\chi(f) \mathcal{F}_{a}}{2f}(\chi\left(  f\right)  -1)\right)  + (\Omega-1) \left(  \frac{F\Omega\tilde{\rho}_{1a}}{\tilde
{\rho}_{a}}-\frac{4\pi A(1-3\Omega)\tilde{\rho}_{a}}{\Omega}-\frac{3+F}{2A}\right)
\label{eqSup3}
\end{equation}
where
\begin{equation}
\tilde{R}=\left[  \frac{\partial\tilde{P}}{\partial r}+\frac{\tilde{\rho
}+\tilde{P}}{1-\frac{2\tilde{m}}{r}}\left(  4\pi r\tilde{P}+\frac{\tilde{m}%
}{r^{2}}\right)  \right]  _{a}\ . \label{rtilde}%
\end{equation}
Equations (\ref{eqSup1}), (\ref{eqSup2}) and (\ref{eqSup3})
conform the \textit{SSE} which coincides with the spherically
symmetric case \cite{AguirreHernandezNunez1994} because that up
to the first order in $\tilde{\alpha},$ the metric functions
$\tilde{m}$ and $\beta$ are found to be independent on the angular
variables. This system may be integrated numerically for any given
radial dependence of the \textit{effective variables,} providing
the total luminosity, a closure relation and a flux factor
$f$\textit{.} The remaining two equations (\ref{TEddrp}) and
(\ref{TEddup}) provide a simple $\theta$-dependence on the
physical variables $\omega_{z},$ and $\mathcal{D}$, i.e.
\begin{equation}
\omega_{z}=\tilde{\alpha}\sin\theta\ \mathcal{Y}\left[  \tilde{m}%
,\beta;\text{their derivatives};r\right]  \quad\text{and}\quad\mathcal{D}%
=\tilde{\alpha}\sin\theta\ \mathcal{Z}\left[  \tilde{m},\beta;\text{ their
derivatives; }r\right]  \label{thetadepomzDe}%
\end{equation}

The restriction due to the junctions conditions for slowly rotating spheroids
(\ref{restricordvarfis}) can be re-written in terms of the effective variables
as
\begin{equation}
2\pi A\frac{\widetilde{\rho}_{a}+\widetilde{P}_{a}}{F}\left(  1+\dot
{A}\right)  =\tilde{v}\left(  u\right)  \alpha\quad\text{and}\quad4\pi
\ A^{2}\ \widetilde{\rho}_{a}\left(  1+\dot{A}\right)  =\tilde{q}\left(
u\right)  \alpha, \label{definieve-quefect}%
\end{equation}
and equation, at least for some models, (\ref{orderparameter}) can be
re-phrased in a very compact form:
\begin{equation}
2AF\Omega\sim1 \label{relaefegefect}%
\end{equation}
Again, this equation becomes very useful when selecting a set of initial
conditions to integrate the \textit{SSE.}

For completeness, we outline here a brief \textit{resumé} of
the HJR method for isotropic slowly rotating radiating fluid
spheres (see \cite{HerreraEtal1994}, for details):

\begin{enumerate}
\item  Take a static interior solution of the Einstein Equations for a fluid
with spherical symmetry, $\rho_{static}=\rho(r)$ and $P_{static}=P(r)$.

\item  Assume that the $r$ dependence of $\tilde{P}$ and $\tilde{\rho}$ are
the same as that of $P_{static}$ and $\rho_{static}$, respectively. Be aware
of the boundary condition:
\begin{equation}
\tilde{P}_{a}=-\omega_{xa}\tilde{\rho}_{a}. \label{condi_borde}%
\end{equation}
and equations (\ref{definieve-quefect}).

\item  With the $r$ dependence of $\tilde{P}$ and $\tilde{\rho}$ and using
(\ref{betayeme}), we get metric elements $\tilde{m}$ and $\beta$ up to some
functions of $u$.

\item  In order to obtain these unknown functions of $u$, we integrate
\textit{SSE}: (\ref{eqSup1}), (\ref{eqSup2}) and (\ref{eqSup3}). The first
two, equations (\ref{eqSup1}) and (\ref{eqSup2}), are model independent, and
the third one, (\ref{eqSup3}), depends of the particular choice of the EoS.

\item  One has four unknown functions of $u$ for the \textit{SSE}. These
functions are: boundary radius $A$, the velocity of the boundary surface
(related to $\Omega$), the total mass $M$ (related to $F$) and the ``total
luminosity'' $L$ Providing one of these functions, a closure relation and the
flux factor $f$, the\textit{\ SSE} can be integrated for any particular set of
initial data a that fulfill equation (\ref{relaefegefect}).

\item  By substituting the result of the integration in the expressions for
$\tilde{m}$ and $\beta$, these metric functions become completely determined.

\item  Again, once we have provided a closure relation and the flux factor $f,$ the set of matter variables, $\omega_{x},$ $\rho,$ $P,$ and $\mathcal{F}$ can be algebraically found for any part of the sphere by using the field equations (\ref{TEdduu})-(\ref{TEddtt}); rotational physical variables, $\omega_{z}$ and $\mathcal{D}$, can be obtained from the remaining significant, two field equations (\ref{TEddrp}) and (\ref{TEddup}). Finally, radiations variables, $\rho_{R}$ and $\mathcal{P}$, emerge from (\ref{anile2}), introducing the \textit{flux factor,} $f,$ the \textit{variable Eddington factor} $\chi$ and any closure relation.
\end{enumerate}

\section{Modelling slowly rotating matter configurations}
\label{Modeling}
In order to explore the influence of the dissipation mechanism and the effect of closure relation on the gravitational collapse of slowly rotating matter configurations, we shall work out three models previously studied for spherical
(nonrotating) cases. We shall work out three different EoS:
\textit{Schwarzschild-like
}\cite{HerreraJimenezRuggeri1980,Tolman1939},
\textit{Tolman IV-like }%
\cite{AguirreHernandezNunez1994,Tolman1939,PatinoRago1983} and
\textit{Tolman VI-like
}\cite{HerreraJimenezRuggeri1980,AguirreHernandezNunez1994,Tolman1939}.

\subsection{The models}
The first family of solutions to be considered is the slowly rotating \textit{Schwarzschild-like} model. In the static limit this model represents an incompressible fluid with constant density. It is the same example presented in ref.
 \cite{HerreraEtal1994} but for the present case we have included the flux \& Eddington factors (\ref{anile2}) and a closure relation from Table \ref{tabla1}. The corresponding effective density and pressure can be written as
\begin{equation}
\tilde{\rho}=k(u)=\frac{3}{8\pi}\frac{1-F}{A^{2}} \quad \text{and} \quad 
\tilde{P}=k(u)\left\{  \frac{3g(u)\left[  1-(8\pi/3)k(u)r^{2}\right]^{1/2}-1}{3-3g(u)\left[  1-(8\pi/3)k(u)r^{2}\right]  ^{1/2}}\right\}  ,
\label{schdefpres}%
\end{equation}
where the function $g(u)$ can be determined from the boundary condition
(\ref{condi_borde}), as
\begin{equation}
g(u)=\frac{3-2\Omega}{\left[  1-(8\pi/3)k(u)a^{2}\right]  ^{1/2}}.
\end{equation}
Third surface equation (\ref{eqSup3}) is
\begin{equation}
\dot{\Omega}=\frac{-\Omega}{1-F}\left[  \frac{3(1-F)^{2}(2\Omega-1)(\Omega
-1)}{2A\Omega}+\frac{\dot{F}}{F}\right]  . \label{oms}%
\end{equation}

The second EoS to be discussed corresponds to the slowly rotating \textit{Tolman-IV-like} model. This model exhibits, in the static limit at the center, the EoS for pure radiation, i.e. $P/{\rho} \sim 1/3$. The effective density and pressure for this case can be expressed as
\begin{equation}
\tilde{\rho}=\frac{1}{8\pi Z(u)}\left\{  \frac{1+3\frac{Z(u)}{W(u)}+3\frac{r^{2}}{W(u)}}{1+2\frac{r^{2}}{Z(u)}}+\frac{1-\frac{r^{2}}{W(u)} }{\left(  1+2\frac{r^{2}}{Z(u)}\right)  ^{2}}\right\}  
\quad \text{and} \quad \tilde{P} = \frac{1-\frac{Z(u)}{W(u)}-3\frac{r^{2}}{W(u)}}{1+2\frac{r^{2}}{Z(u)}}; 
\label{effectivTO4}
\end{equation}
where
\[
Z(u)  =-\frac{A^{2}\left[  7(1-F)+2\Omega(F-2)-\eta\right] }{2(F-1)(2\Omega+3)},\quad
\text{and} \quad
W(u)   =\frac{-A^{2}\left(  F(1+2\Omega)-1+\eta\right)  }{2\left[F(2-3\Omega)+F^{2}(6\Omega-5)+1+(F-1)\eta\right]  };
\]
with
\[
\eta=\sqrt{1+F(22-20\Omega)+F^{2}(4\Omega^{2}+2\Omega-23)}.
\]{equation}

For this model, the third surface equation can be written as
\begin{equation}
\dot{\Omega}=\Theta\dot{A}+\Phi\dot{F}+\Gamma
\end{equation}
where the expression for the coefficients $\Theta,$ $\Phi,$ and $\Gamma$ in terms of the \textit{Surface Variables} and their derivatives (i.e. $A,F,\Omega,\dot{A},$and $\dot{F}$) are sketched in the Appendix \ref{ThirdEq}.

The third family of models is inspired on \textit{Tolman VI }static solution, which approaches the one of a highly relativistic Fermi gas, with the corresponding adiabatic exponent of $4/3$. For this case we have
\begin{equation}
\tilde{\rho}=\frac{3h(u)}{r^{2}}=\frac{1}{8\pi r^{2}}(1-F)\qquad
\text{and}\qquad\tilde{P}=\frac{h(u)}{r^{2}}\frac{\left(  1-9d(u)r\right)
}{\left(  1-d(u)r\right)  }.\label{effectivTO6}%
\end{equation}
As before, the function $d(u)$ is determined from the equation (\ref{condi_borde}), thus, we obtain
\begin{equation}
d(u)=\frac{1}{3}\frac{4\Omega-1}{A(4\Omega-3)}.
\end{equation}
We find that the third surface equation is
\begin{equation}
\dot{\Omega}=\frac{4F^{2}(F-1)\Omega^{3}(2\Omega^{2}-5\Omega+2)+2\dot
{F}A\Omega(\Omega-1)^{2}+F(3-2F-F^{2})\Omega^{2}-F(F-1)^{2}(1-3\Omega)}{2AF(F-1)(\Omega-1)^{2}}.
\label{tet}%
\end{equation}

\subsection{General considerations}
For all the models, we have selected a set of initial conditions and physical parameters that resemble, as much as possible, interesting astrophysical scenarios. We have chosen
\begin{itemize}
\item  Typical values of these conditions that resembles young neutron stars. Notice that because the coupling restriction for slow rotation assumption (\ref{orderparameter}) (or its equivalent in the adimensional variables
(\ref{relaefegefect})) the initial radius of the configuration becomes $4$ times greater than the typical neutron star radius.

\item  In our simulations, we have imposed that the energy conditions for perfect fluid be satisfied. In addition, the restrictions $-1<\omega_{x},\omega_{z}<1$ and $r>2\tilde{m}\left(  u,r\right)  $ at any shell within the matter configuration are also fulfilled.

\item  As it was pointed out at the end of the preceding section, the evolution of one of the variable at the surface (boundary radius $A$, the velocity of the boundary surface (related to $\Omega$), the total mass $M$
(related to $F$) or the ``total luminosity'' $L$) has to be provided. For the present simulation the evolution of the luminosity profile, $L(u)$, is given as a Gaussian pulse centered at $u=u_{p}$
\begin{equation}
-\dot{M}=L=\frac{\Delta M_{rad}}{\lambda\sqrt{2\pi}}\exp\frac{1}{2}\left(\frac{u-u_{p}}{\lambda}\right)^{2}, 
\label{eq:mpunto}%
\end{equation}
where $\lambda$ is the width of the pulse and $\Delta M_{rad}$ is the total mass lost in the process.

\item  Throughout the simulations equations (\ref{definiefe-ge}) (or equivalently (\ref{orderparameter}) or (\ref{relaefegefect})) are constantly checked in order to verify the validity of the approximation.
\end{itemize}

\subsection{Modeling radiation transfer scenarios}
\label{modelingscenarios}
\begin{figure}
\begin{center}
\includegraphics[width=5in]{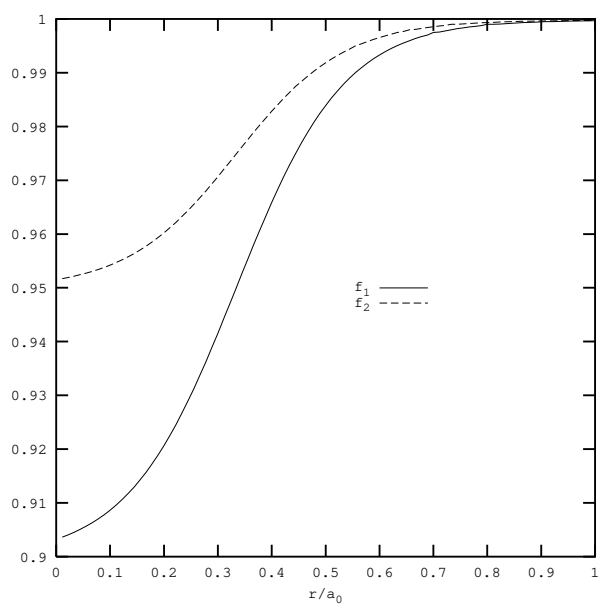}%
\caption{Variable Flux factor $f=f\left(  x=\frac{r}{m(0) }\right)  =\dfrac{\left(
f_{surface}+e^{-\zeta(x_{t}-x)}f_{core}\right) }{\left( 1+e^{-\zeta(x_{t}-x)}\right)  }$where $f_{1}~=~f_{core}~=~0.902$ which is considered the\ Schwarzschild-like and Tolman IV-like models, and changing\ $f_{2}~=~f_{core}~=~0.952$ for the Tolman VI-like models.}
\label{f10}%
\end{center}
\end{figure}

We would like to explore how dissipation affects the dynamics of these three types of\ slowly rotating matter distributions. For each of the above EoS and  the several closure relations listed in Table \ref{tabla1}, we shall work out simulations with:
\begin{enumerate}
\item \textbf{Matter configurations with constant \textit{flux factor} }$f.$ We study several radiation transfer environments ranging from the collapse of opaque matter distribution where $f~=~0.426$ (close to a \textit{diffusion regime}) to more transparent matter configuration where $f~=~0.930$ where the radiation transport mechanism is described near the \textit{free streaming out limit approximation.}

\item \textbf{Matter configuration with variable \textit{flux factor}
}$f=f(r).$ For this case we study the effect of a variable flux profile as
\begin{equation} 
f=f\left(x=\frac{r}{m(0)  } \right) = \dfrac{{e}^{-\zeta(x_{t}-x)}f_{core}+f_{surface}}{1+{e}^{-\zeta(x_{t}-x)}}
\label{varibleflxfact}%
\end{equation}
on the orbital velocity at the equator. We have defined $f_{core}$ as the flux factor at the inner core and $f_{surface}$ the flux factor at the surface of the distribution. The parameters $x_{t}~=~ {r_{t}}/{m\left(  0\right)  }$ represents the cutoff region where the transition of the dissipation mechanism takes
place and $\zeta$ regulates how sharp or smooth is the transition
between the flux factors at two regions (see Figure \ref{f10}).
The idea with this variable flux factor is to allow configurations
with more opaque matter at the inner core and more transparent
mass shells at the outer mantle. In many Astrophysical scenarios,
radiation diffuses out from a central opaque region
($\chi(f\nolinebreak =\nolinebreak 0)\nolinebreak =\nolinebreak
{1}/{3}$) to the transparent boundary ($\chi(f\nolinebreak
=\nolinebreak 1)\nolinebreak =\nolinebreak 1$).
\end{enumerate}

\subsection{Parameters and astrophysical  scenarios}
Concerning the luminosity (equation (\ref{eq:mpunto})), all our models have been been simulated with 
\begin{equation}
\Delta M_{rad}=2.00\times10^{-11}\,M(0),\qquad\lambda=0.74\times
10^{-3}\mathrm{s},\qquad t_{p}=1.48\times10^{-3}\mathrm{s.} \label{pulset}%
\end{equation}

For the modelling with the Lorentz-Eddington closure relation and the three above mentioned EoS we have set initial conditions and parameters to have the following values:
\begin{center}%
\begin{tabular}
[c]{|rl||rl|}\hline
$m(0)=$ 		& $1${$.0M_{\odot}$} 	& $\alpha=$ 								& ${10^{-3}}$ \\
$A(0)=$ 		& $6,000$ 			& {$\Rightarrow a\left(  0\right)  \equiv r_{Surface}=$}& $44,500\,m$ \\
$\Omega(0)=$  & ${0.999}$ 			& $\Rightarrow${$\omega_{a}\left(  0\right)  =$} 	&$-0.00101\,c$\\\hline\hline
\end{tabular}
\end{center}
The Lorentz-Eddington closure relation was initially proposed by C.D. Levermore in the early 80's \cite{Levermore1984} on the basis of geometrical considerations, for the case of stationary medium. Later, it has been reobtained by other authors from different perspectives, i.e., thermodynamical point of view with maximum
entropy principles \cite{AnilePennisiSammartino1991} and Information Theory with the energy flux taken as a constraint \cite{Dominguez1997}.

For all other closure relations listed in Table \ref{tabla1}, the set of initial conditions and parameters for the modelling performed with
the above seed EoS are:

\begin{center}%
\begin{tabular}
[c]{|rl||rl|}\hline
$m(0)=$ & $1${$.0M_{\odot}$} & $\alpha=$ & ${10^{-3}}$\\
$A(0)=$ & $41,104$ & {$\Rightarrow a\left(  0\right)  \equiv r_{Surface}=$} &
$36,500\,\,m$\\
$\Omega(0)=$ & ${0.999}$ & $\Rightarrow${$\omega_{a}\left(  0\right)  =$} &
$-0.00101\,c$\\\hline\hline
\end{tabular}
\end{center}

Concerning the flux factor, $f$, we have considered two different scenarios for the three equations of state and for the six closure relations.  First, for simplicity, we are going to consider the case $f = const$. The second, and more realistic, scenario, described by equation (\ref{varibleflxfact}) and displayed in Figure \ref{f10}, has the corresponding the set of parameters:
\begin{equation}
f_{surface}=1;\qquad f_{core}=0.902;\qquad\zeta=10\qquad\text{and}\qquad
r_{t}=14,833m
\end{equation}

\section{Summary of results, comments and conclusions}
\label{Conclusions}
We have extended a previous work \cite{HerreraEtal1994} to study the collapse of a radiating, slowly rotating self-gravitating relativistic configuration by introducing the \textit{flux factor}, the \textit{Variable Eddington Factor} (equations (\ref{anile2})) and a closure relation (Table \ref{tabla1}).  With this extension, in principle, it is possible to implement the seminumerical approach to simulate a variety of radiation transport mechanism within a slowly rotating radiating matter distribution that can be used as evaluation testbeds for emerging full-numerical environments.

Using this seminumerical approach it has been explored the influence of the closure relation on the dynamics of collapsing slowly rotating relativistic bodies. The results we found can be summarized as follows. 
  First, it has been obtained that, for all closure relations listed in Table \ref{tabla1}, the junction conditions imply that total diffusion regime can not be attained at the surface of the configuration. This result is exact and general, independent of the EoS and valid for spherical and axisymmetric matter configurations. 

Secondly (also related to the junction conditions), it is evident from equation (\ref{eccentgeneral}), that the eccentricity at the surface of radiating configurations (up to first order in $\tilde{\alpha}$) is greater for models near the diffusion limit approximation than for those in the free streaming out limit. Again, this result is EoS independent and is present for all closure relations we have studied.

\begin{figure}
\begin{center}
\includegraphics[width=4.5in]{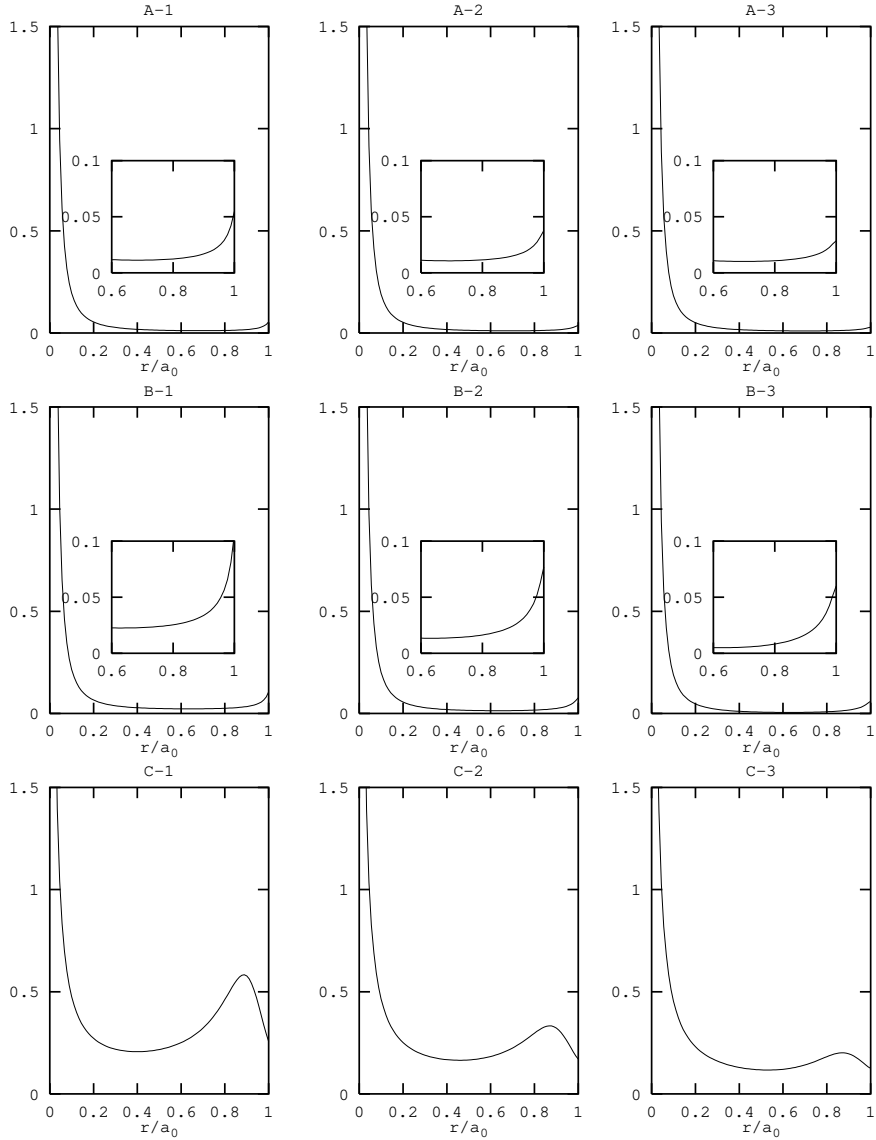}%
\caption{Profiles of orbital velocity $\omega_{z} \times10^{-6}c,$ for the Schwarzschild-like (plates A-1 thought
A-3), Tolman IV-like (plates B-1 thought B-3) and Tolman VI-like (plates C-1 thought C-3) at three distinct times\newline $u=10,30,50.$ The profiles in each plate correspond to a variable Flux factor\newline $f_{LE}~=f_{LE}~\left(  x=\frac{r}{m(0)  }\right)  =\dfrac{\left(  e^{-\zeta(x_{t}-x)}f_{LE}{}_{surface} + f_{core}\right)  }{\left(  1+e^{-\zeta(x_{t}-x)}\right)  }.$}%
\label{f11}%
\end{center}
\end{figure}

\begin{figure}
\begin{center}
\includegraphics[width=4.5in]{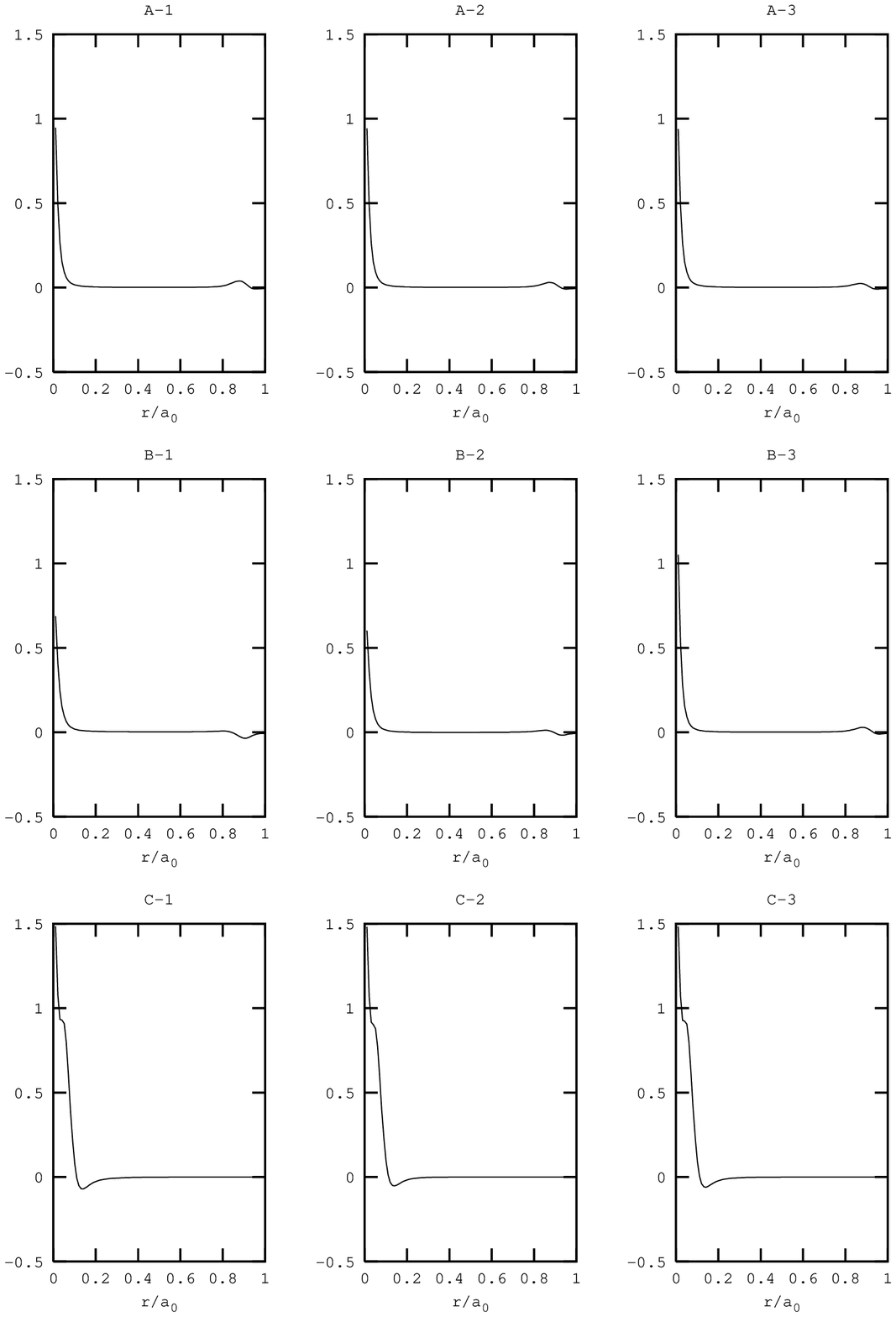}%
\caption{Profiles of orbital velocity $\omega_{z} \times10^{-6}c,$ corresponding \textit{Minerbo} closure relation.
They are represented in A-1 thought A-3 (Schwarzschild-like); B-1 thought B-3 (Tolman IV-like) and plates C-1 thought C-3 (Tolman VI-like) at three distinct times $u=10,30,50.$ The profiles in each plate correspond to a variable Flux factor $f_{Mi}~=f_{Mi}~\left(  x=\frac{r}{m(0)  }\right) =\dfrac{\left(f_{Mi}+e^{-\zeta(x_{t}-x)}f_{core}\right)  }{\left( 1+e^{-\zeta(x_{t}-x)}\right)  }.$}%
\label{f15}%
\end{center}
\end{figure}

\begin{figure}
\begin{center}
\includegraphics[width=4.5in]{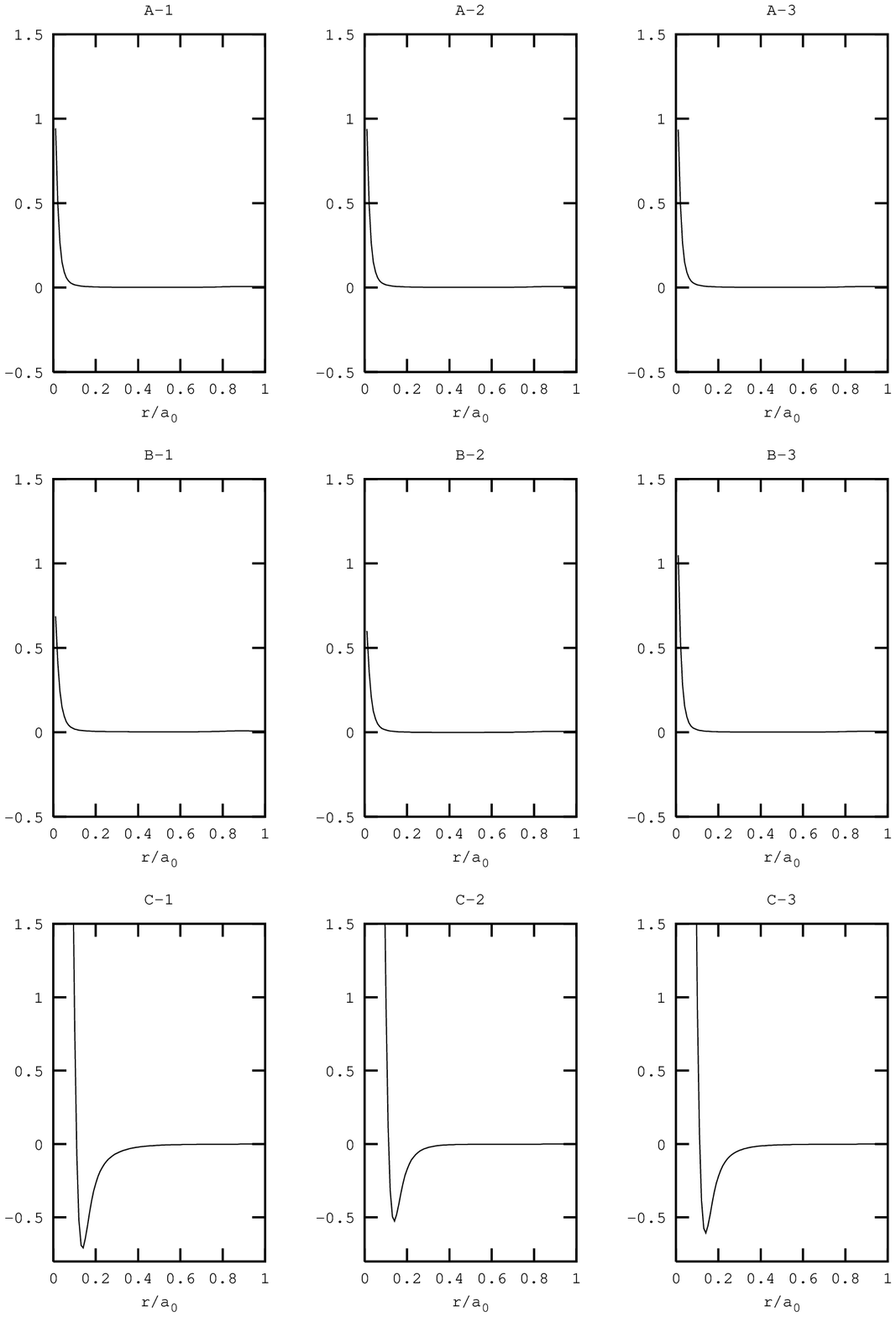}%
\caption{Profiles of orbital velocity $\omega_{z} \times10^{-6}c,$ corresponding \textit{Janka }(\textit{Monte
Carlo}) closure relation. They are represented in A-1 thought A-3 Schwarzschild-like); B-1 thought B-3 (Tolman IV-like) and plates C-1 thought C-3 (Tolman VI-like) at three distinct times $u=10,30,50.$ The profiles in each plate correspond to a variable Flux factor $f_{MC}~=f_{MC}~\left(  x=\frac{r}{m\left(  0\right)
}\right)  =\dfrac {\left( f_{MC}{}_{surface}+e^{-\zeta(x_{t}-x)}f_{core}\right)  }{\left( 1+e^{-\zeta(x_{t}-x)}\right)  }.$}%
\label{f16}%
\end{center}
\end{figure}

\begin{figure}
\begin{center}
\includegraphics[width=4.5in]{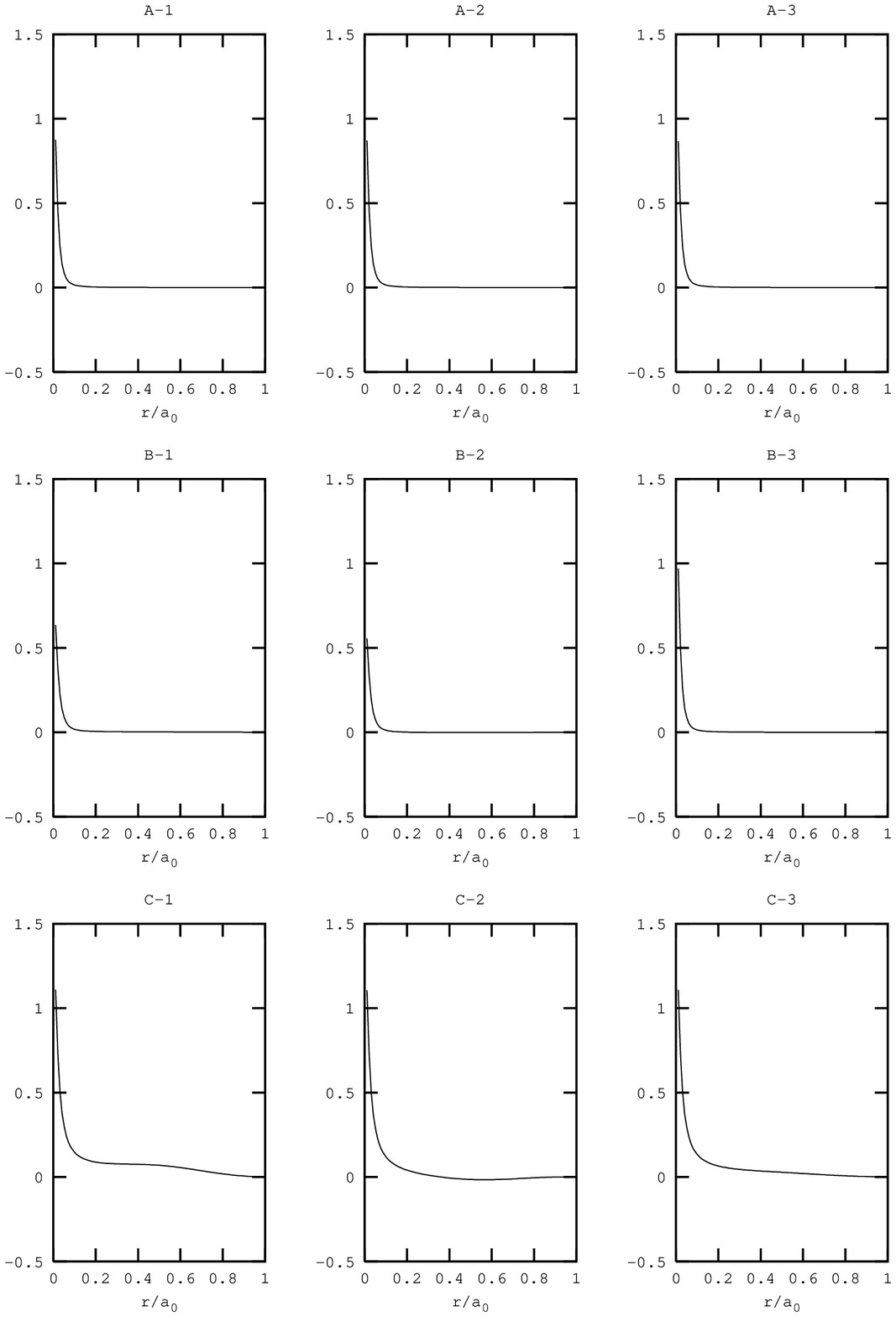}%
\caption{Profiles of orbital velocity $\omega_{z}\times10^{-6}c,$ corresponding \textit{Maximum Packing} closure
relation. They are represented in A-1 thought A-3 (Schwarzschild-like); B-1 thought B-3 (Tolman IV-like) and plates
C-1 thought C-3 (Tolman VI-like) at three distinct times $u=10,30,50.$ The profiles in each plate correspond to a variable Flux factor $f_{MP}~=f_{MP}~\left(  x=\frac{r}{m\left(  0\right)  }\right)
=\dfrac{\left(f_{MP}{}_{surface}+e^{-\zeta(x_{t}-x)}f_{core}\right)  }{\left(1+e^{-\zeta(x_{t}-x)}\right)  }.$}%
\label{f17}%
\end{center}
\end{figure}

The third result is the significant influence of the closure relations on the dynamic of physical variables. For the more realistic scenario concerning a variable flux factor (equation (\ref{varibleflxfact})) and for the first order approximation in rotating parameters, we have found qualitatively different behavior for the profiles of the tangential velocity $\omega_{z}$. As it can be appreciated from figures \ref{f11}-\ref{f17} we obtain models having outer layers rotating faster than the inner ones or experimenting counter rotation for some mass shells surrounding the nucleus.  

\begin{figure}
\begin{center}
\includegraphics[width=4.5in]{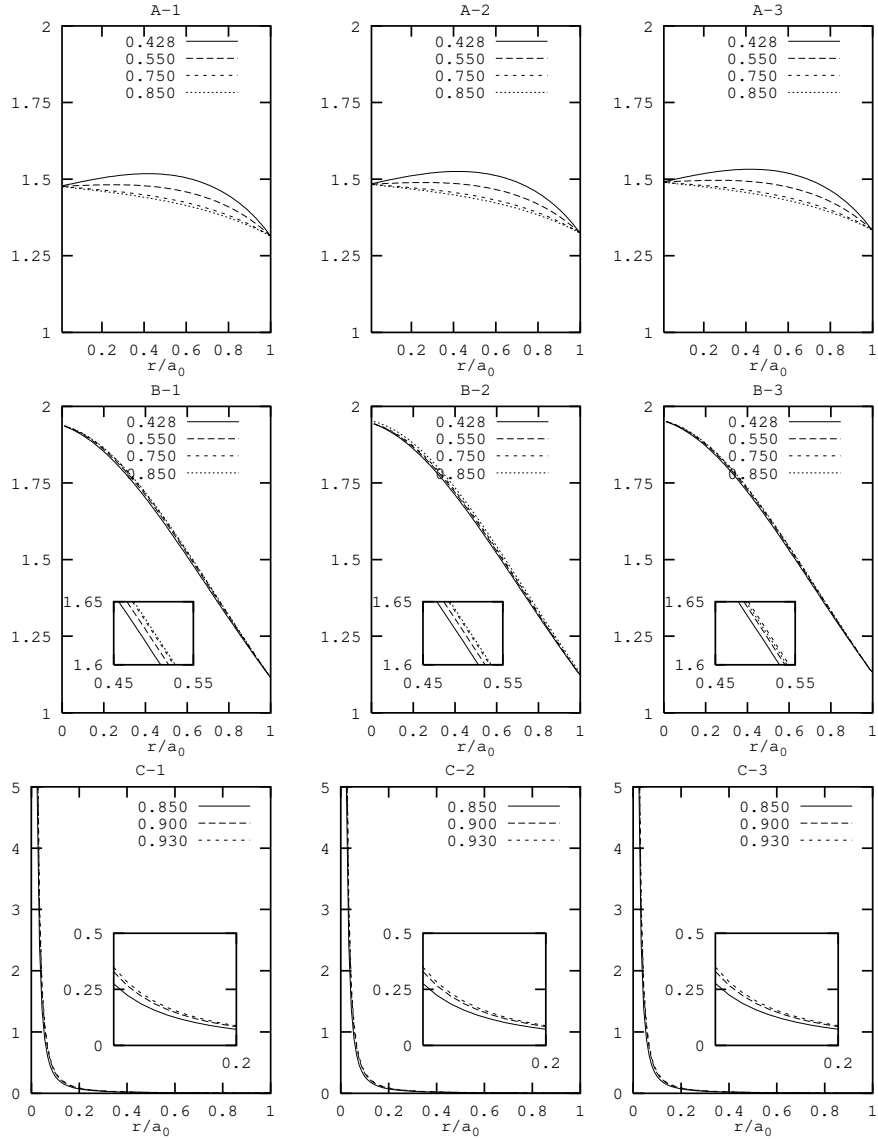}%
\caption{Profiles of hydrodynamic density, $\rho\times10^{14}$ gr/cm$^{3}$, corresponding to Schwarzschild-like and for Tolman IV-like are represented in plates A-1 thought A-3 and B-1 thought B-3, respectively. Tolman VI-like models are displayed in plates C-1 thought C-3 as $\rho\times10^{16}$ gr/cm$^{3}$. The various flux
factors are $f_{LE}~=~0.426,0.550,0.750,0.850$ for Schwarzschild-like and the Tolman IV-like models and $f_{LE}
~=~0.850,0.900,0.930$ for the Tolman VI-like. The retarded timesdisplayed are $u=10,30,50.$}
\label{f2}%
\end{center}
\end{figure}

\begin{figure}
\begin{center}
\includegraphics[width=4.5in]{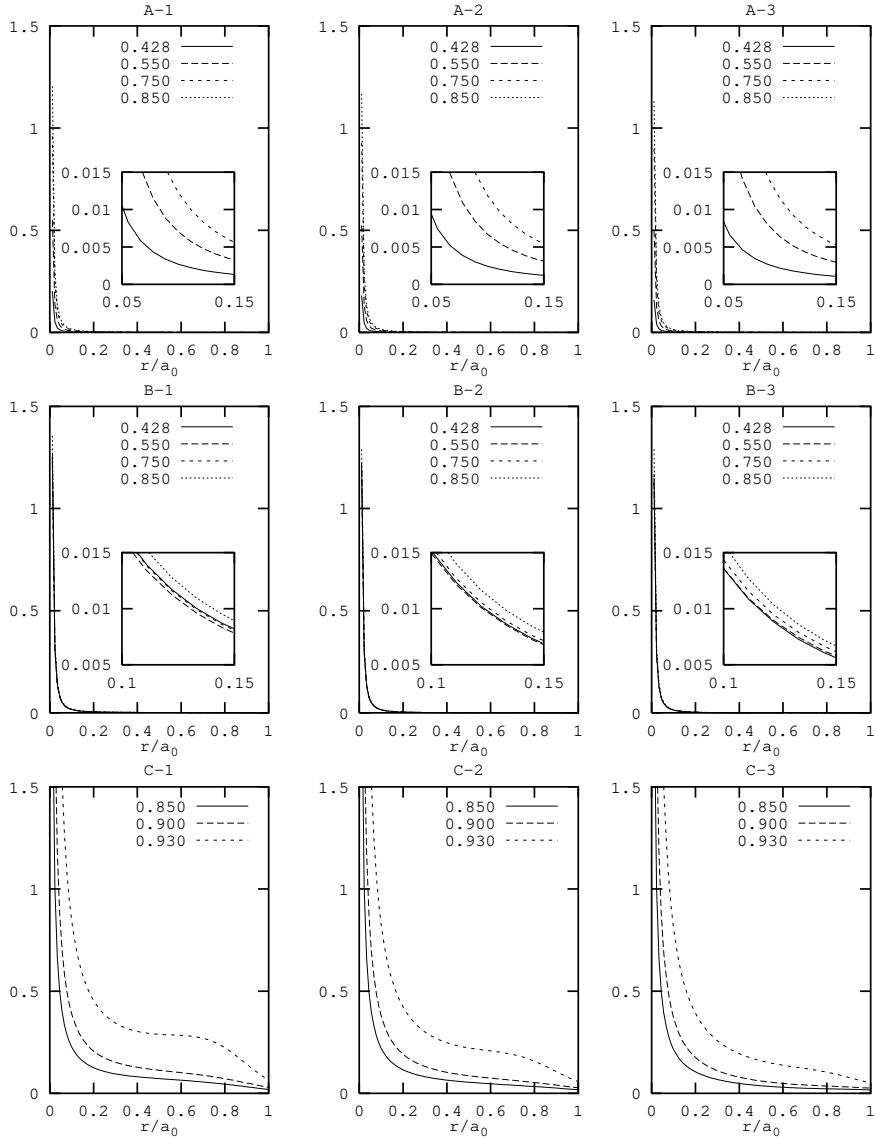}
\caption{Profiles of orbital velocity $\omega_{z}\times10^{-6}c,$ corresponding to Schwarzschild-like for Tolman IV-like and Tolman VI-like are respresented in plates (A-1 thought A-3), (B-1 thought B-3) and (C-1 thought C-3), respectively. The various constant flux factors are $f_{LE}~=~0.426,0.550,0.750,0.850$ for Schwarzschild-like and the Tolman IV-like models and $f_{LE}~=~0.850,0.900,0.930$ for the Tolman VI-like. The retarded times displayed are $u=10,30,50.$}%
\label{f8}%
\end{center}
\end{figure}

\begin{figure}
\begin{center}
\includegraphics[width=4.5in]{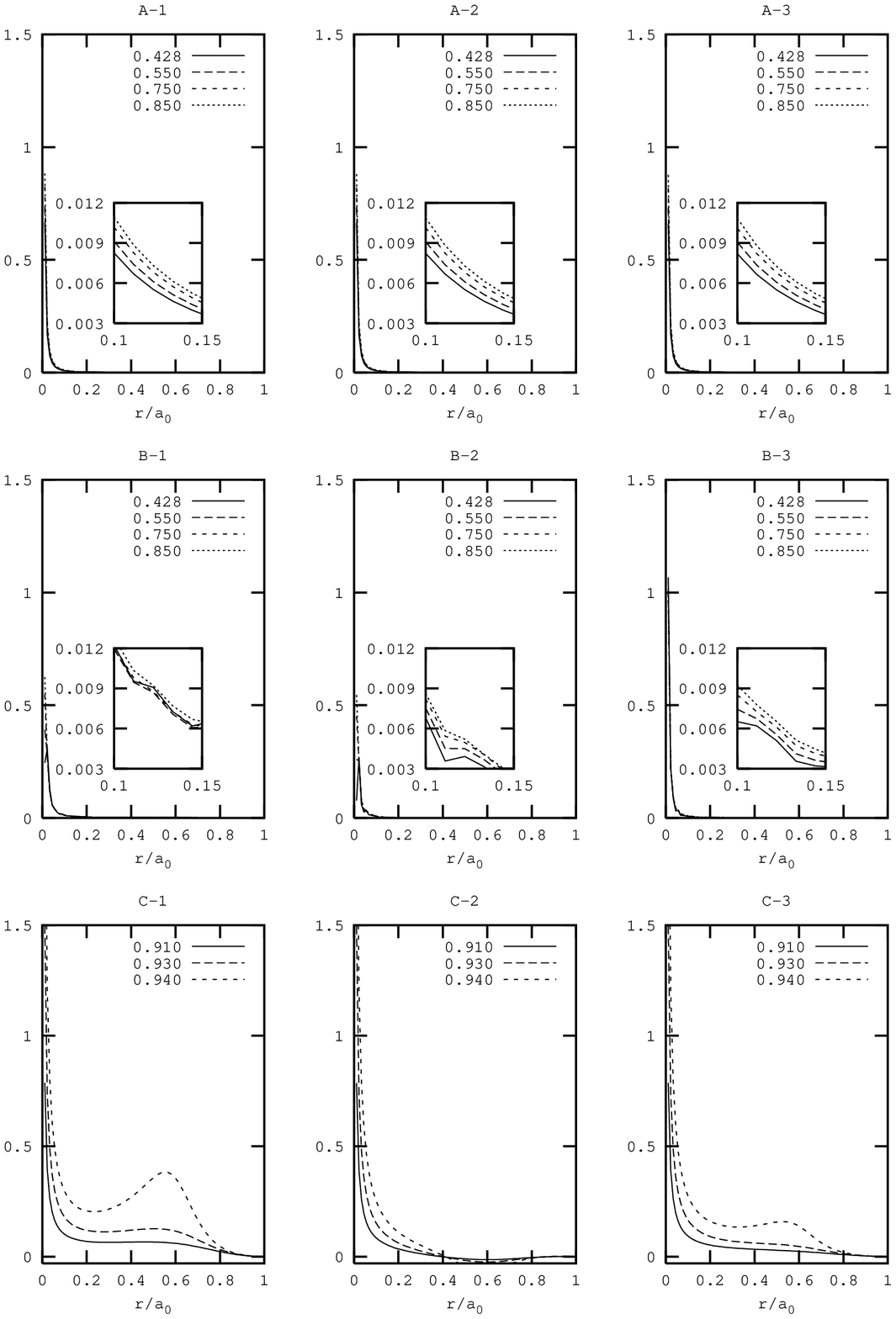}%
\caption{Profiles of orbital velocity corresponding to \textit{Minerbo} closure relation. They are represented in A-1
thought A-3 (Schwarzschild-like); B-1 thought B-3 (Tolman IV-like) and plates C-1 thought C-3 (Tolman VI-like) with $\omega_{z} \times10^{-5}c.$ The constant flux factors are $f_{Mi}~=~0.428,0.550,0.750,0.850$ for Schwarzschild-like and the Tolman IV-like models and $f_{Mi}~=~0.910,0.930,0.940$ for the Tolman VI-like with $\omega_{z}\times10^{-4}c$. The retarded times displayed are $u=10,30,50.$}
\label{f12}%
\end{center}
\end{figure}

\begin{figure}
\begin{center}
\includegraphics[width=4.5in]{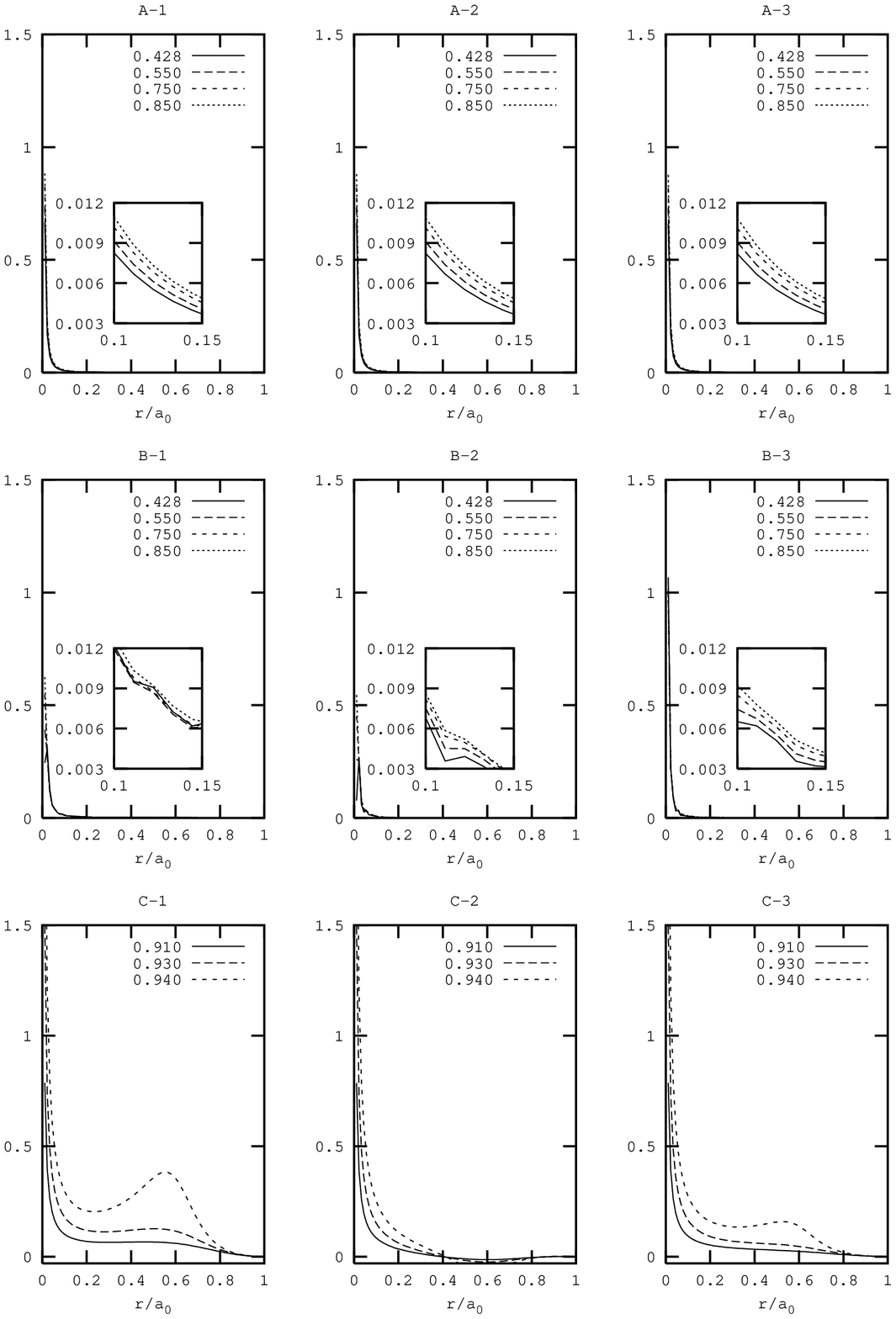}
\caption{Profiles of orbital velocity corresponding to \textit{Janka }(\textit{Monte Carlo}) closure relation. They are
represented in A-1 thought A-3 (Schwarzschild-like); B-1 thought B-3 (Tolman IV-like) and plates C-1 thought C-3 (Tolman VI-like) with $\omega_{z}$ $\times10^{-5}c.$ The constant flux factors are $f_{MC}~=~0.428,0.550,0.750,0.850$ for Schwarzschild-like and the Tolman IV-like models and $f_{MC}~=~0.910,0.930,0.940$ for the Tolman VI-like with $\omega_{z} \times10^{-4}c$. The retarded times displayed are $u=10,30,50.$}
\label{f13}
\end{center}
\end{figure}

\begin{figure}
\begin{center}
\includegraphics[width=4.5in]{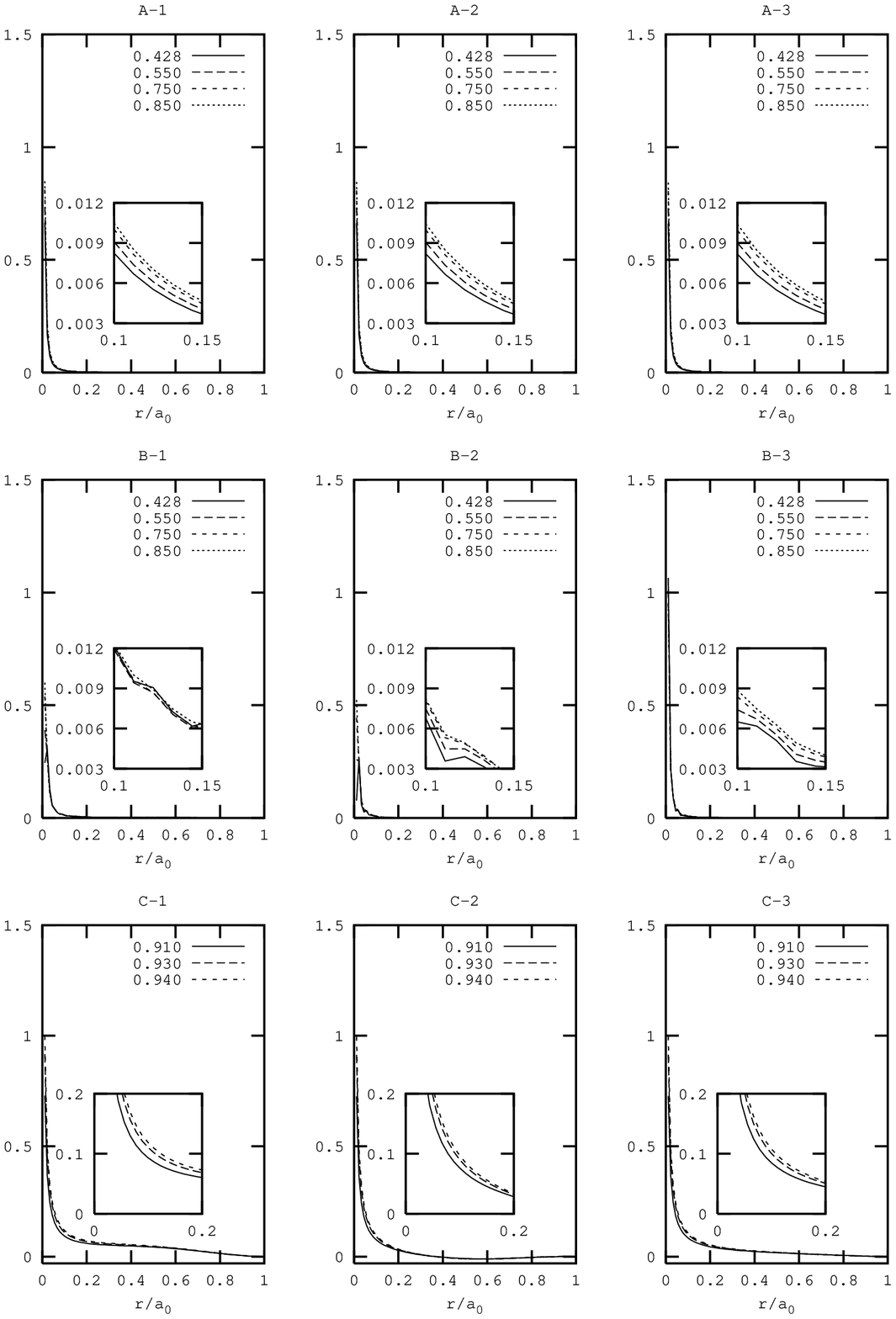}%
\caption{Profiles of orbital velocity corresponding to \textit{Maximum Packing} closure relation. They are represented in A-1 thought A-3 (Schwarzschild-like); B-1 thought B-3 (Tolman  IV-like) and plates C-1 thought C-3 (Tolman VI-like) with $\omega_{z}$ $\times10^{-5}c.$ The constant flux factors are  $f_{MP}~=~0.428,0.550,0.750,0.850$ for Schwarzschild-like and the Tolman IV-like models and $f_{MP}~=~0.910,0.930,0.940$ for the
Tolman VI-like with $\omega_{z}$ $\times10^{-4}c$. The retarded times displayed are $u=10,30,50.$}%
\label{f14}%
\end{center}
\end{figure}

Finally some other results emerge from our modeling. Concerning newtonian theory, it is clear that the gravitational effects of rotation are purely relativistic and as it can be understood recalling that the newtonian parameter measuring the ``strength'' of rotation is not linear in the angular velocity but proportional to the square of it. In this work, rotation is considered in the slow approximation limit, i.e., we dealing with situations where the tangential velocity of every fluid element is much less than the speed of light and the centrifugal forces are little compared with the gravitational ones.  Despite that the physical rationale to study the gravitational effects of rotation in the slow rotation approximation seems to be justified, this assumption appears to be very restrictive when junction conditions are considered. At least for the modeling performed, we have found that combinations of significant physical variables (equations (\ref{restricordvarfis}) and (\ref{orderparameter}) among others) are forced to maintain the first order approximation and, consequently, the most plausible astrophysical scenario we have obtained within this limit (and EoS), are relativistic rotating matter configurations surrounded by an extended  ``atmospheres''. This is more evident from Tolman VI-like model (plates C-1, C2 and C-3 in figure  \ref{f2}) where, in spite of its singularity, it could represent an object having hydrodynamic densities $\rho \sim (10-20) \rho_{0}$ at a core $0 < r \lessapprox 10$ Kms. with a thinner matter distribution ($\rho\lessapprox10^{16}$ gr/cm$^{3}$)\ prolonged to the outer mantle $10$ Kms. $\lessapprox r \lessapprox40$ Kms. Schwarzschild-like and Tolman IV-like models (plates A-1, through B-3 in the same figures) provide the same range for hydrodynamic densities, but for configurations having radius $4$ times greater than typical ones for neutron stars.   We have also found that all the obtained models experiment differential rotation, i.e. the core rotates much faster than the envelope. But more important than this, is the effect of the dissipation on the orbital velocity. From figures \ref{f8}, \ref{f12}, \ref{f13} and \ref{f14} it can be appreciated that the more diffusive the model is, the slower it rotates. This effect seems to be independent of closure relation and the EoS considered.  Notice that we have mainly displayed here figures related to the tangential velocity $\omega_{z}$. For the complete set of figures for all physical variables corresponding every EoS in each scenario  and details of some of the calculation, the interested reader is referred to the website \newline
 \url{http://webdelprofesor.ula.ve/ciencias/nunez/EddintonFactor/OtherFiguresVEddingtonFac.html}.

\section{Acknowledgments}
We gratefully acknowledge the financial support of the Consejo de Desarrollo Científico Humanístico y Tecnológico de la Universidad de Los Andes (CDCHT-ULA) under project C-1009-00-05-A, and to the Fondo Nacional de Investigaciones Científicas y Tecnológicas (FONACIT) under projects S1-2000000820 and F-2002000426. Two of us (F.A. and T.S.) have been benefited by the computational infrastructure and consulting support of the Centro Nacional de
Cálculo Científico, Universidad de Los Andes (\textsc{CeCalCULA}).

\section{Appendix}

\subsection{Einstein equations for slowly rotating matter configuration}
\label{EinsteinEquations}
Denoting differentiation with respect to $u,$ $r$ and $\theta$ are denoted by subscripts $0$,$1$ and $2$, respectively, Einstein equations for slowly rotating matter configuration can be written as:
\begin{itemize}
\item $8\pi T_{uu}=G_{uu}:$%
\begin{align}
&  \left(  1-\frac{2\overset{\thicksim}{m}}{r}\right)  \frac{8\pi r^{2}%
}{\omega_{x}^{2}-1}\left[  \rho+\rho_{R}-\mathcal{F}+\left(  2\omega
_{x}+1\right)  \mathcal{F}+\omega_{x}^{2}\left(  P+\mathcal{P}\right)  \right]
\nonumber\\
& \label{Tuu}\\
&  =\qquad2\left(  1-\frac{2\overset{\thicksim}{m}}{r}\right)  \overset
{\thicksim}{m}_{1}-\frac{\overset{\thicksim}{m}_{22}}{r}-\frac{\overset
{\thicksim}{m_{2}}}{r}\cot\theta-2e^{-2\beta}\overset{\thicksim}{m_{0}%
}+3\left(  1-\frac{2\overset{\thicksim}{m}}{r}\right)  \beta_{2}%
-\frac{6\overset{\thicksim}{m}_{2}}{r}\beta_{2}\ ,\nonumber
\end{align}

\item $8\pi T_{ur}=G_{ur}:$%
\begin{equation}
\frac{8\pi r^{2}}{1+\omega_{x}}\left[  \rho+\rho_{R}-\mathcal{F}-\omega
_{x}\left(  P+\mathcal{P}-\mathcal{F}\right)  \right]  =2\overset{\thicksim
}{m}_{1}-\beta_{22}-\beta_{2}\cot\left(  \theta\right)  -\beta_{2}^{2}\ ,
\label{Tur}%
\end{equation}

\item $8\pi T_{rr}=G_{rr}:$%
\begin{equation}
2\pi r\left(  1-\frac{2\overset{\thicksim}{m}}{r}\right)  ^{-1}\left(
\frac{1-\omega_{x}}{1+\omega_{x}}\right)  \left[  \rho+P+\rho_{R}%
+\mathcal{P}-2\mathcal{F}\right]  =\beta_{1}\ , \label{Trr}%
\end{equation}

\item $8\pi T_{\theta\theta}=G_{\theta\theta}:$%
\begin{align}
4\pi r^{2}\left[  2P-\mathcal{P}+\rho_{R}\right]   &  =\qquad2\beta_{2}%
\cot\theta-\overset{\thicksim}{m}_{11}r-2e^{-2\beta}r^{2}\beta_{01}-6\beta
_{1}\overset{\thicksim}{m}_{1}r\nonumber\\
& \label{Ttt}\\
+3\beta_{1}r+\beta_{2}^{2}  &  +\left(  1-\frac{2\overset{\thicksim}{m}}%
{r}\right)  \left(  4\beta_{1}^{2}r+2\beta_{11}r-\beta_{1}\right)
r\ ,\nonumber
\end{align}

\item $8\pi T_{u\theta}=G_{u\theta}:$%
\begin{equation}
0=\left(  1-\frac{2\overset{\thicksim}{m}}{r}\right)  \left(  r\beta
_{21}+4r\beta_{1}\beta_{2}-\beta_{2}\right)  -\overset{\thicksim}{m}%
_{21}-2\beta_{2}\overset{\thicksim}{m}_{1}+\frac{\overset{\thicksim}{m}_{2}%
}{r}-e^{-2\beta}r\beta_{02}-4\beta_{1}\overset{\thicksim}{m}_{2}\ ,
\label{Tut}%
\end{equation}

\item $8\pi T_{r\theta}=G_{r\theta}:$%
\begin{equation}
0=\beta_{21}-\frac{2}{r}\beta_{2}\ , \label{Trt}%
\end{equation}

\item $8\pi T_{\theta\phi}=G_{\theta\phi}:$%
\begin{align}
0 &  =\left(  1-\frac{2\overset{\thicksim}{m}}{r}\right)  r^{2}\left(
4\beta_{2}\beta_{1}-\beta_{21}\right)  +\overset{\thicksim}{m}_{2}\left(
1-4\beta_{1}r\right)  -r\overset{\thicksim}{m}_{21}\nonumber\\
& \label{Ttp}\\
&  \qquad-2\beta_{2}\left(  \overset{\thicksim}{m}_{1}r-\overset{\thicksim}%
{m}\right)  -e^{-2\beta}r^{2}\left(  \beta_{21}-\beta_{02}\right)  -2\beta
_{2}e^{-2\beta}r^{2}\left(  \beta_{1}-\beta_{0}\right)  \ ,\nonumber
\end{align}

\item $8\pi T_{r\phi}=G_{r\phi}:$%
\begin{align}
&  8\pi r\left\{  -\frac{r}{2\sin\theta}\left(  \frac{1-\omega_{x}}%
{1+\omega_{x}}\right)  ^{\frac{1}{2}}\left(  1-\frac{2\overset{\thicksim}{m}%
}{r}\right)  ^{-\frac{1}{2}}\left(  2\mathcal{F}+\mathcal{P}-3\rho_{R}\right)
\mathcal{D}\right.  \nonumber\\
&  -\overset{\thicksim}{\alpha}\left(  1-\frac{2\overset{\thicksim}{m}}%
{r}\right)  ^{-1}\left(  \frac{\omega_{x}-1}{\omega_{x}+1}\right)  \left(
\rho+P+\rho_{R}+\mathcal{P}-2\mathcal{F}\right)  -\overset{\thicksim}{\alpha
}\frac{e^{2\beta}}{\omega_{x}+1}\left[  \rho+\rho_{R}-\mathcal{F}-\omega
_{x}\left(  P+\mathcal{P}-\mathcal{F}\right)  \right]  \nonumber\\
&  +\frac{r}{2\sin\theta}\left(  1-\frac{2\overset{\thicksim}{m}}{r}\right)
^{-\frac{1}{2}}\frac{\omega_{z}}{\omega_{x}\left(  \omega_{x}+1\right)
}\left[  -2\omega_{x}\left(  P+\mathcal{P}+\mathcal{F}\left(  \frac{1}%
{f}-1\right)  \right)  +\right.  \label{Trp}\\
&  \left.  \left.  \left(  3\mathcal{P-}\rho_{R}-2\mathcal{F}\right)  \left(
\omega_{x}+1-\sqrt{1-\omega_{x}^{2}}\right)  \right]  \right\}  \nonumber\\
&  =\qquad\overset{\thicksim}{\alpha}\left[  2r\left(  \beta_{1}+\beta
_{0}\right)  -1+r^{2}\left(  \beta_{11}-\beta_{01}\right)  +e^{2\beta}\left(
1-2\overset{\thicksim}{m}_{1}\right)  +e^{2\beta}\beta_{2}\left(  \beta
_{2}-3\cot\theta\right)  +e^{2\beta}\beta_{22}\right]  \ ,\nonumber
\end{align}

\item $8\pi T_{u\phi}=G_{u\phi}:$%
\begin{align}
&  8\pi r^{2}\left\{  \frac{r}{2\sin\theta\sqrt{1-\omega_{x}^{2}}}\left(
1-\frac{2\overset{\thicksim}{m}}{r}\right)  ^{\frac{1}{2}}\left(  2\omega
_{x}\mathcal{F}-\mathcal{P}+3\rho_{R}\right)  \mathcal{D}+\right.  \nonumber\\
&  \frac{\overset{\thicksim}{\alpha}}{\omega_{x}+1}\left[  2\left(  \rho
+\rho_{R}-\mathcal{F}\right)  -\omega_{x}\left(  P+\mathcal{P}-\mathcal{F}%
\right)  \right]  +\frac{\overset{\thicksim}{\alpha}e^{2\beta}}{\omega_{x}%
^{2}-1}\left(  1-\frac{2\overset{\thicksim}{m}}{r}\right)  \left[  \rho
+\rho_{R}-\mathcal{F}+\left(  \omega_{x}+1\right)  ^{2}\mathcal{F}\right.
\nonumber\\
&  \left.  +\omega_{x}^{2}\left(  P+\mathcal{P}-\mathcal{F}\right)  \right]
+\frac{r\omega_{z}}{2\omega_{x}\sin\theta\left(  1-\omega_{x}^{2}\right)
}\left[  2\omega_{x}\left(  \rho+P+\rho_{R}+\mathcal{P}-2\mathcal{F}\right)
+2\mathcal{F}\left(  \omega_{x}+1\right)  ^{2}\right.  \nonumber\\
&  \left.  \left.  +\sqrt{1-\omega_{x}^{2}}\left[  2\mathcal{F}+\omega
_{x}\left(  3\mathcal{P}-\rho_{R}\right)  \right]  \right]  \right\}
\nonumber\\
&  =\qquad\overset{\thicksim}{\alpha}\left\{  \left(  1-\frac{2\overset
{\thicksim}{m}}{r}\right)  \left[  r^{2}\left(  4\beta_{0}\beta_{1}-4\beta
_{1}^{2}+\beta_{01}\right)  -\left(  \beta_{0}-3\beta_{1}\right)  r-1\right]
\right.  \label{Tup}\\
&  \qquad+r\left[  \left(  \beta_{0}-3\beta_{1}\right)  \left(  1-2\overset
{\thicksim}{m}_{1}\right)  -\overset{\thicksim}{m}_{01}+\overset{\thicksim}%
{m}_{11}-\frac{\overset{\thicksim}{m}_{0}}{r}\left(  4\beta_{1}r-3\right)
\right]  +e^{2\beta}\left(  1-\frac{2\overset{\thicksim}{m}}{r}\right)
\left(  1-2\overset{\thicksim}{m}_{1}\right)  \nonumber\\
&  \qquad+e^{-2\beta}r^{2}\left(  \beta_{01}-\beta_{00}\right)  -e^{2\beta
}\left(  1-\frac{2\overset{\thicksim}{m}}{r}\right)  \left(  3\beta_{2}%
^{2}+6\beta_{2}\cot\theta+e^{-2\beta}\beta_{11}r^{2}\right)  \nonumber\\
&  \qquad\left.  +\frac{e^{2\beta}}{r}\left(  \overset{\thicksim}{m}%
_{22}+6\beta_{2}\overset{\thicksim}{m}_{2}+3\overset{\thicksim}{m}_{2}\right)
-\beta_{2}\left(  \beta_{2}-3\cot\theta\right)  -\beta_{22}\right\}
\ .\nonumber
\end{align}
\end{itemize}

\subsection{Einstein equations in terms of the flux and Eddington factors }
\label{EinsteinFluxFactor}
Einstein equations for slowly rotating matter configuration can be written in terms of the flux and Eddington factor  as:
\begin{itemize}
\item $8\pi T_{uu}=G_{uu}:$%
\begin{align}
&  \left(  1-\frac{2\overset{\thicksim}{m}}{r}\right)  \frac{8\pi r^{2}%
}{\omega_{x}^{2}-1}\left[  \rho+\omega_{x}^{2}P+\left(  \frac{1}{f}%
+2\omega_{x}+\omega_{x}^{2}\frac{\chi}{f}\right)  \mathcal{F}\right]
\nonumber\\
& \label{TEdduu}\\
&  =2\left(  1-\frac{2\overset{\thicksim}{m}}{r}\right)  \overset{\thicksim
}{m}_{1}-\frac{\overset{\thicksim}{m}_{22}}{r}-\frac{\overset{\thicksim}%
{m_{2}}}{r}\cot\theta-2e^{-2\beta}\overset{\thicksim}{m_{0}}+3\left(
1-\frac{2\overset{\thicksim}{m}}{r}\right)  \beta_{2}-\frac{6\overset
{\thicksim}{m}_{2}}{r}\beta_{2}\ ,\nonumber
\end{align}

\item $8\pi T_{ur}=G_{ur}:$%
\begin{equation}
\frac{8\pi r^{2}}{1+\omega_{x}}\left[  \rho-\omega_{x}P+\left(  \frac{1}%
{f}-1+\omega_{x}\left(  1-\frac{\chi}{f}\right)  \right)  \mathcal{F}\right]
=2\overset{\thicksim}{m}_{1}-\beta_{22}-\beta_{2}\cot\left(  \theta\right)
-\beta_{2}^{2}\ , \label{TEddur}%
\end{equation}

\item $8\pi T_{rr}=G_{rr}:$%
\begin{equation}
2\pi r\left(  1-\frac{2\overset{\thicksim}{m}}{r}\right)  ^{-1}\left(
\frac{1-\omega_{x}}{1+\omega_{x}}\right)  \left[  \rho+P+\left(  \frac{1}%
{f}+\frac{\chi}{f}-2\right)  \mathcal{F}\right]  =\beta_{1}\ , \label{TEddrr}%
\end{equation}

\item $8\pi T_{\theta\theta}=G_{\theta\theta}:$%
\begin{align}
4\pi r^{2}\left[  2P+\frac{1}{f}\left(  1-\chi\right)  \mathcal{F}\right]   &
=2\beta_{2}\cot\theta-\overset{\thicksim}{m}_{11}r-2e^{-2\beta}r^{2}\beta
_{01}-6\beta_{1}\overset{\thicksim}{m}_{1}r\nonumber\\
& \label{TEddtt}\\
&  +3\beta_{1}r+\beta_{2}^{2}+\left(  1-\frac{2\overset{\thicksim}{m}}%
{r}\right)  \left(  4\beta_{1}^{2}r+2\beta_{11}r-\beta_{1}\right)
r\ ,\nonumber
\end{align}

\item $8\pi T_{u\theta}=G_{u\theta}:$%
\begin{equation}
0=\left(  1-\frac{2\overset{\thicksim}{m}}{r}\right)  \left(  r\beta
_{21}+4r\beta_{1}\beta_{2}-\beta_{2}\right)  -\overset{\thicksim}{m}%
_{21}-2\beta_{2}\overset{\thicksim}{m}_{1}+\frac{\overset{\thicksim}{m}_{2}%
}{r}-e^{-2\beta}r\beta_{02}-4\beta_{1}\overset{\thicksim}{m}_{2}\ ,
\label{TEddut}%
\end{equation}

\item $8\pi T_{r\theta}=G_{r\theta}:$%
\begin{equation}
0=\beta_{21}-\frac{2}{r}\beta_{2}\ , \label{TEddrt}%
\end{equation}

\item $8\pi T_{\theta\phi}=G_{\theta\phi}:$%
\begin{align}
0 &  =\left(  1-\frac{2\overset{\thicksim}{m}}{r}\right)  r^{2}\left(
4\beta_{2}\beta_{1}-\beta_{21}\right)  +\overset{\thicksim}{m}_{2}\left(
1-4\beta_{1}r\right)  -r\overset{\thicksim}{m}_{21}\nonumber\\
& \label{TEddtp}\\
&  \qquad-2\beta_{2}\left(  \overset{\thicksim}{m}_{1}r-\overset{\thicksim}%
{m}\right)  -e^{-2\beta}r^{2}\left(  \beta_{21}-\beta_{02}\right)  -2\beta
_{2}e^{-2\beta}r^{2}\left(  \beta_{1}-\beta_{0}\right)  \ ,\nonumber
\end{align}

\item $8\pi T_{r\phi}=G_{r\phi}:$%
\begin{align}
&  8\pi r\left\{  -\frac{r}{2\sin\theta}\left(  \frac{1-\omega_{x}}%
{1+\omega_{x}}\right)  ^{\frac{1}{2}}\left(  1-\frac{2\overset{\thicksim}{m}%
}{r}\right)  ^{-\frac{1}{2}}\left(  2+\frac{\chi}{f}-\frac{3}{f}\right)
\mathcal{FD}\right.  \nonumber\\
&  \qquad\qquad-\overset{\thicksim}{\alpha}\left(  1-\frac{2\overset
{\thicksim}{m}}{r}\right)  ^{-1}\left(  \frac{\omega_{x}-1}{\omega_{x}%
+1}\right)  \left(  \left(  \frac{1}{f}+\frac{\chi}{f}-2\right)
\mathcal{F}+\rho+P\right)  \nonumber\\
&  \qquad\qquad-\overset{\thicksim}{\alpha}\frac{e^{2\beta}}{\omega_{x}%
+1}\left[  \left(  \frac{1}{f}-1-\omega_{x}\left(  \frac{\chi}{f}-1\right)
\right)  \mathcal{F}+\rho-\omega_{x}P\right]  \label{TEddrp}\\
&  \qquad\qquad+\left.  \frac{r}{2\sin\theta}\left(  1-\frac{2\overset
{\thicksim}{m}}{r}\right)  ^{-\frac{1}{2}}\frac{\omega_{z}}{\omega_{x}\left(
\omega_{x}+1\right)  }\left[  P+\left(  \omega_{x}\left(  \chi-3\right)
+3\chi+\left(  3\chi-1-2f\right)  \sqrt{\left(  1-\omega_{x}^{2}\right)
}\right)  \frac{\mathcal{F}}{f}\right]  \right\}  \nonumber\\
&  =\overset{\thicksim}{\alpha}\left[  2r\left(  \beta_{1}+\beta_{0}\right)
-1+r^{2}\left(  \beta_{11}-\beta_{01}\right)  +e^{2\beta}\left(
1-2\overset{\thicksim}{m}_{1}\right)  +e^{2\beta}\beta_{2}\left(  \beta
_{2}-3\cot\theta\right)  +e^{2\beta}\beta_{22}\right]  \ ,\nonumber
\end{align}

\item $8\pi T_{u\phi}=G_{u\phi}:$%
\begin{align}
&  8\pi r^{2}\left\{  \frac{r}{2\sin\theta\sqrt{1-\omega_{x}^{2}}}\left(
1-\frac{2\overset{\thicksim}{m}}{r}\right)  ^{\frac{1}{2}}\left(  2\omega
_{x}-\frac{\chi}{f}+\frac{3}{f}\right)  \mathcal{FD}\right.  +\nonumber\\
&  \qquad\qquad+\frac{\overset{\thicksim}{\alpha}}{\omega_{x}+1}\left[
2\rho-\omega_{x}P+\left(  \frac{2}{f}-2-\omega_{x}\left(  \frac{\chi}%
{f}-1\right)  \right)  \mathcal{F}\right]  +\frac{\overset{\thicksim}{\alpha
}e^{2\beta}}{\omega_{x}^{2}-1}\left(  1-\frac{2\overset{\thicksim}{m}}%
{r}\right)  \left[  \frac{\rho f+\omega_{x}^{2}Pf}{f}\right.  +\nonumber\\
&  \qquad\qquad+\left.  \frac{1+2\omega_{x}f+\omega_{x}^{2}\chi}{f}%
\mathcal{F}\right]  +\frac{r\omega_{z}}{\omega_{x}\sin\theta\left(
1-\omega_{x}^{2}\right)  }\left[  \frac{\omega_{x}+\omega_{x}\chi+f\omega
_{x}^{2}+f}{f}\mathcal{F}+\frac{\omega_{x}\rho f+\omega_{x}Pf}{f}\right.
\nonumber\\
&  \qquad\qquad\left.  \left.  +\sqrt{1-\omega_{x}^{2}}\left(  1+\frac
{\omega_{x}}{2}\left(  3\frac{\chi}{f}-\frac{1}{f}\right)  \right)
\mathcal{F}\right]  \right\}  \nonumber\\
&  =\overset{\thicksim}{\alpha}\left\{  \left(  1-\frac{2\overset{\thicksim
}{m}}{r}\right)  \left[  r^{2}\left(  4\beta_{0}\beta_{1}-4\beta_{1}^{2}%
+\beta_{01}\right)  -\left(  \beta_{0}-3\beta_{1}\right)  r-1\right]  \right.
+\label{TEddup}\\
&  \qquad\qquad+r\left[  \left(  \beta_{0}-3\beta_{1}\right)  \left(
1-2\overset{\thicksim}{m}_{1}\right)  -\overset{\thicksim}{m}_{01}%
+\overset{\thicksim}{m}_{11}-\frac{\overset{\thicksim}{m}_{0}}{r}\left(
4\beta_{1}r-3\right)  \right]  +e^{2\beta}\left(  1-\frac{2\overset{\thicksim
}{m}}{r}\right)  \left(  1-2\overset{\thicksim}{m}_{1}\right)  +\nonumber\\
&  \qquad\qquad+e^{-2\beta}r^{2}\left(  \beta_{01}-\beta_{00}\right)
-e^{2\beta}\left(  1-\frac{2\overset{\thicksim}{m}}{r}\right)  \left(
3\beta_{2}^{2}+6\beta_{2}\cot\theta+e^{-2\beta}\beta_{11}r^{2}\right)
+\nonumber\\
&  \qquad\qquad+\left.  \frac{e^{2\beta}}{r}\left(  \overset{\thicksim}%
{m}_{22}+6\beta_{2}\overset{\thicksim}{m}_{2}+3\overset{\thicksim}{m}%
_{2}\right)  -\beta_{2}\left(  \beta_{2}-3\cot\theta\right)  -\beta
_{22}\right\}  \ .\nonumber
\end{align}
\end{itemize}

\subsection{The third surface equation}
\label{ThirdEq}
As we have stated in Section \ref{Modeling}, the third surface equation for
the slowly rotating \textit{Tolman-IV-like} model can be written as
\[
\dot{\Omega}=\Theta\dot{A}+\Phi\dot{F}+\Gamma
\]
where the expression for the coefficients $\Theta,$ $\Phi,$ and $\Gamma$ in
terms of the \textit{Surface Variables} are:
\begin{equation*}
\Theta  =\frac{\Omega}{2A\mathcal{U}}\left\{  2\Omega^{2}F\left( 12\Omega^{2}F+22F+\mathcal{H}-30\right)  -\Omega\left(  149F^{2} +146F-3\mathcal{H}F+3-\mathcal{H}\right)  \right.  +\left.  92F^{2}-88F-4\right\}
\end{equation*}%
and
\begin{align*}
\Phi  = & -\frac{1}{2}\frac{\Omega}{F\mathcal{Q}}\left\{  2\Omega^{2}%
F^{2}(6\Omega F-4\Omega+22F+\mathcal{H}-47)+3+\mathcal{H}\right. \\
&  \left.  \qquad +\Omega(44\Omega F-149F^{3}-3F^{2}\mathcal{H}+257F^{2}%
-106F-2)+F(92F-157F-\mathcal{H}+62)\right\}
\end{align*}
with
\begin{align*}
\mathcal{U}    = &\Omega F\left[  4\Omega F\left(  \Omega+4\right)
-22\Omega-63F+62\right]  +\Omega+2F\left(  23F-22\right)  -2,\\
& \\
\mathcal{Q}    = &\Omega F\left[  4\Omega F\left(  \Omega F-\Omega+4F\right)
-\left(  38F-22\right)  \Omega-\left(  63F-125\right)  F-61\right] -\Omega+F\left(  46F^{2}-90F+42\right)  +2\\
&  \text{and}\\
\mathcal{H}    = &\sqrt{1+22F-20\Omega F-23F^{2}+20\Omega F^{2}+4\Omega
^{2}F^{2}};
\end{align*}
finally
\[
\Gamma=\sum_{k=0}^{5}c_{k}\Omega^{(k)}%
\]
where
\[
c_{0}=96\pi A^{3}F(2F-1)(F-1)
\]%
\begin{align*}
c_{1}   = & 2AF((-8F\mathcal{T}L+6F^{2}+4\mathcal{T}L-9+3F)\mathcal{H}+108F-267F^{2}+150F^{3}+9-4\mathcal{T}L(3F-2F^{2}-1)\\
& \qquad  \qquad+\pi A^{2}(-576F^{2}+864F-288))
\end{align*}%
\begin{align*}
c_{2}    = & (6F^{3}+24AF^{3}+60AF+12F-3-15F^{2}-84AF^{2})\mathcal{H} +\pi A^{3}F(2112F^{2}-3168F+1056)-588AF^{2}-48AF\\
&  \qquad-888AF^{4}+1524AF^{3}-3-16\mathcal{T}LF^{2}A(2F-1) +F(66F^{3}+117F-21-159F^{2})
\end{align*}%
\begin{align*}
c_{3}   = & 6(F-1)(-16AF^{2}+2F+8AF-1)\mathcal{H} + 6(F-1)(80AF^{3}-34F^{3}+49F^{2}-48AF^{2}-14F\\
&  \qquad-96\pi A^{3}F(2F+1)-4AF-1)
\end{align*}%
\[
c_{4}=-12F(2F-1)(F-1)\mathcal{H}+12F(2F-1)(F-1)(5F+8AF-6) \quad \text{and} \quad c_{5}=24F^{2}(+1+2F^{2}-3F)
\]
with
\[
\mathcal{T}=\frac{1}{2f}-\frac{3}{2}\frac{\chi}{f}+1\quad\text{and}\quad
L=4\pi A^{2}(2\Omega-1)\epsilon.
\]

\pagebreak %

%
%
\pagebreak %

%
%

\pagebreak %
%
%

\pagebreak %
%
%

\pagebreak %
%
%

\pagebreak %
%
%
\clearpage

%
%

\clearpage
%
%

\pagebreak %
%
%

\pagebreak %
%
%

\pagebreak %
%
%


\newpage
\begin{thebibliography}{99}
\bibitem{Demianski1985}Demianski, M. (1985), \textit{Relativistic Astrophysics,} in International Series in Natural Philosophy, Vol 110, Edited by \textit{D. Ter Haar}, (Pergamon Press, Oxford).

\bibitem{ShapiroTeukolsky1983}Shapiro, S.L. and Teukolsky, S.A. (1983), \textit{Black Holes, White Dwarfs and Neutron Stars}, (John Willey, New York).

\bibitem{KippenhahnWeigert1990}Kippenhahn, R. and Weigert, A. (1990), \textit{Stellar Structure and Evolution}, (Springer Verlag, New York).

\bibitem{Glendening2000}Glendenning, N. K. (2000), \textit{Compact Stars} (Springer Verlag, New York).

\bibitem{BruennDeNiscoMezzacappa2001}Bruenn, S.W., De Nisco, K.R., and Mezzacappa, A. (2001) \textit{Astrophys J.}, \textbf{560}, 326. [Online article]: cited on 23 Jan 2001 ,\url{\ http://arXiv.org/abs/astro-ph/0101400}

\bibitem{HartleThorne1969}Hartle, J.B., and Thorne, K.S. (1969), \textit{Astrophys. J}., \textbf{158}, 719.

\bibitem{Stergioulas2003}Stergioulas, N. (2003), ``Rotating Stars in Relativity'', \textit{Living Rev. Relativity} \textbf{6}. [Online article]: cited on 25 Dec 2000 \url{http://www.livingreviews.org/lrr-2003-3/} also [Online article]: cited on 10 Feb 2003 in \url{http://arXiv.org/abs/gr-qc/0302034}

\bibitem{Font2003}Font, J. (2003) ``Numerical Hydrodynamics in General Relativity'', \textit{Living Rev. Relativity} \textbf{6}, 4.[Online article]: cited on 25 Dec 2000 \url{http://www.livingreviews.org/lrr-2003-4/} also in
[Online article]: cited on 12 May 2003 \url{http://arXiv.org/abs/gr-qc/0003101}

\bibitem{Lorimer2001}Lorimer, D.R. (2001) ``Binary and Millisecond Pulsars at the New Millenium'',\textit{\ Living Rev. Relativity} \textbf{4}, 5. [Online article]: cited on 15 Aug 2001 \url{http://www.livingreviews.org/lrr-2001-5}

\bibitem{ChineaGonzalez-Romero1993}Chinea, F.J. and González-Romero, L.M. (1993), \textit{Rotating Objects and Relativistic Physics}. Lecture Notes in Physics \textbf{423}. Springer Verlag.

\bibitem{MankoMielkeSanabria2000}Manko, V.S., Mielke, E.W., and Sanabria-Gómez, J.D.\ (2000) \textit{Phys. Rev. D}, \textbf{61}, 081501-1.[Online article]: cited on 23 April 2002, \url{http://www.arxiv.org/abs/gr-qc/0001081} Manko, V.S., Sanabria-Gómez, J.D., and Manko, O.V., (2000) \textit{Phys. Rev. D}, \textbf{62}, 044048-1.

\bibitem{BertiStergioulas2004}Berti, E., and Stergioulas, N., (2004) \textit{Mon. Not. R. Astr. Soc}, \textbf{350,} 1416. [Online article]: cited on 16 Feb 2004, \url{http://www.arxiv.org/abs/gr-qc/0310061}

\bibitem{BertiEtal2004}Berti, E., White, F., Maniopoulou, A. and Bruni, M. (2005) \textit{Mon.Not.Roy.Astron.Soc.} \textbf{358 }, 923. [Online article]: cited on 29 May 2004 \url{http://arxiv.org/abs/gr-qc/0405146}

\bibitem{PachonRuedaSanabria2006} Pachón, L. A., Rueda, J.A  and Sanabria-Gómez, J.D. (2006) \textit{Phys.Rev.} \textbf{D73} 104038. [Online article]: cited on 13 Jun 2006 \url{http://arxiv.org/abs/gr-qc/0606060}

\bibitem{SibgatullinSunyaev2000}Sibgatullin, N.R., and Sunyaev, R.A. (2000)\textit{ Astron. Lett.}, \textbf{26}, 699. [Online article]: cited on 23 April 2002, \url{http://www.arxiv.org/abs/astro-ph/0011253}

\bibitem{SiebelEtall2003}Siebel, F., Font, J.A., M\"{u}ller, E., and Papadopoulos, P. (2003) \textit{Phys. Rev. D} \textbf{67}, 124018. [Online article]: cited on 7 February 2003, \url{http://www.arxiv.org/abs/gr-qc/0301127}

\bibitem{DimmelmeierFontMueller2002}Dimmelmeier, H., Font, J. A., and Mueller, E. (2002) \textit{Astron. Astrophys.} \textbf{393}, 523 [Online article]: cited on 17 Apr 2002, \url{http://arXiv.org/abs/astro-ph/0204289}

\bibitem{NASAGC}The NASA Grand Challenge Project is described at \url{http://wugrav.wustl.edu/Relativ/nsgc.html}

\bibitem{GRAstro3D}The code GR Astro and its documentation can be found at \url{http://wugrav.wustl.edu/Codes/GR3D}

\bibitem{Cactus}For information see \url{http://www.cactuscode.org}

\bibitem{WalderEtal2005}Walder, R., Burrows, A., Ott, E., Livne, V., Lichtenstadt, I.  and Jarrah, M. (2005) \textit{Astrophys J.}, \textbf{626}, 317. [Online
article]: cited on 23 February 2005 ,\url{http://arXiv.org/abs/astro-ph/0412187}

\bibitem{LiebendorferEtal2001}Liebend\"{o}rfer, M., Mezzacappa, A.,
Thielemann, F., \textit{et al}.(2001), \textit{Phys. Rev. D}, \textbf{63}, 3004. [Online article]: cited on 28 Jun 2000,\url{http://arXiv.org/abs/astro-ph/0006418}

\bibitem{LiebendorferEtal2002}Liebend\"{o}rfer, M, O. E. B., Mezzacappa A.
\textit{et al (2002) A Finite Difference Representation of Neutrino Radiation Hydrodynamics for Spherically Symmetric General Relativistic Supernova Simulations.} [Online article]: cited on 24 Oct 2003, \url{http://arXiv.org/abs/astro-ph/0207036}

\bibitem{RamppJanka2002}Rampp M. and Janka H.-Th. (2002) \textit{Astron.Astrophys,.}\textbf{396}, 361\textit{.} [Online article]: cited on 7 Mar 2002, \url{http://arXiv.org/abs/astro-ph/0203101}

\bibitem{HerreraJimenezRuggeri1980}Herrera, L., Jiménez, J. and Ruggeri, G. (1980) \textit{Phys. Rev.} \textbf{D22,} 2305.

\bibitem{HerreraNunez1990}{Herrera, L. and Núñez, L. A. (1990)\textit{\ Fund. Cosmic Phys.} \textbf{14}, 235.}

\bibitem{HernandezNunezPercoco1998}Hernández, H., Núñez, L.A., and Percoco, U. (1999), \textit{Class. Quantum Grav}, \textbf{16}, 871. [Online article]: cited on 5 June, 1998, \url{http://arXiv.org/abs/gr-qc/9806029}

\bibitem{BarretoMartinezRodriguez2002}Barreto,W., Martínez, H. and Rodríguez B., (2002) \textit{Ap. Space Sc. }\textbf{282} 581.

\bibitem{HerreraEtal2002}Herrera L., Barreto W., Di Prisco A., and Santos N.O. (2002) \textit{Phys. Rev.} \textbf{D65,} 104004 [Online article]: cited on 14 Feb 2002, \url{http://arXiv.org/abs/gr-qc/0202051}

\bibitem{Hartle1967}Hartle, J. B. (1967) \textit{Astrophys. J.}, \textbf{150}, 1005.

\bibitem{HartleThorne1968}Hartle, J. B., and Thorne, K. S. (1968) \textit{Astrophys. J.}, \textbf{153}, 807.

\bibitem{HerreraEtal1994}Herrera, L., Melfo, A., Núñez, L.A. and Patiño, A. (1994) \textit{Astrophys. J.} 421, 677.

\bibitem{Levermore1984}Levermore C. D. (1984) \textit{J. Quant. Spectrosc. Radiat. Transfer,} \textbf{31}, 149.

\bibitem{Dominguez1997}Domínguez Cascante, R. (1997) \textit{Jour. of Phys A,} \textbf{30} 7707. [Online article]: cited on 9 Sep 1997, \url{http://www.arXiv.org/abs/cond-mat/9709104}

\bibitem{PonsIbanezMiralles2000}Pons, J.A. Ibañez, J. M$^{\mathrm{a}}$ and Miralles J. A. (2000) \textit{Mon. Not. R. Astr. Soc}, \textbf{317}, 550. [Online article]: cited on 15 May 2000, \url{http://arXiv.org/abs/astro-ph/0005310}

\bibitem{SmitVandenHornBludman2000}Smit, J.M.,Van den Horn, L.J., and Bludman, S.A. (2000) \textit{Astron. Astrophys.} \textbf{356}, 559.

\bibitem{OstrikerBodenheimerLyndenBell1966}Ostriker, J. P., Bodenheimer, P, and Lynden-Bell, D., (1966) \textit{Phys. Rev Lett.}, \textbf{17}, 816.

\bibitem{BaumgartShapiroShibata2000}Baumgarte, T. W., Shapiro, S. L., and Shibata, M., (2000), \textit{Astrophys. J},, \textbf{528}, L29. [Online article]: cited on 1 Nov 1999,\url{ http://arXiv.org/abs/astro-ph/9910565}

\bibitem{LyfordBaumgarteShapiro2002}Lyford, N.D., Baumgarte, T.W. and Shapiro,
S. L. (2003) \textit{Astrophys.J.,} \textbf{583} 410 [Online article]: cited
on 3 Oct 2002, \url{http://arXiv.org/abs/gr-qc/0210012}

\bibitem{Bondi1964}Bondi, H., (1964) \textit{Proc. Roy. Soc. London,}
\textbf{281,} 39.

\bibitem{CarmeliCaye1977}Carmeli, M. and Kaye, M., (1977) \textit{Ann. Phys.
(N.Y.),}\textbf{\ 103,} 97.

\bibitem{GonzalezHerreraJimenez1979}González,C., Herrera, L. and
Jiménez J., (1979) \textit{J. Math.Phys}., \textbf{20}, 836.

\bibitem{KramerHahner1995}Kramer, D. and Hahner, U., (1995) \textit{Class. Quantum. Grav.},\textbf{\ 12}, 2287.

\bibitem{HerreraJimenez1982}Herrera, L. and Jiménez, J., (1982) \textit{\ J. Math. Phys}.,\textbf{\ 23,} 2339.

\bibitem{BondiVandenburgMetzner1962}{Bondi, H., Van der Burg, M. G. J. and
Metzner, A. W. K., (1962) \textit{Proc. R. Soc. London,} \textbf{A269}, 21.}

\bibitem{HerreraSantos1997}Herrera, L. and Santos N.O.
(1997),\textit{\ Physics Reports,} \textbf{286}, 53.

\bibitem{Lindquist1966}{Lindquist R. W. (1966)\textit{\ Ann. Phys.}  \textbf{\ 37,} 487.}

\bibitem{MihalasMihalas1984}{Mihalas, D. and Weibel Mihalas, B. (1984) \textit{Foundations of Radiation Hydrodynamics} (Oxford University Press).}

\bibitem{RezzollaMiller1994}Rezzolla L. and Miller J. (1994) \textit{Class. Quantum Grav.} \textbf{11,} 1815.

\bibitem{AndersonSpiegel1972}Anderson J.L. and Spiegel E.A. (1972) \textit{Astrophys. J.} \textbf{171}, 127.

\bibitem{AliRomano1994}Ali G. and Romano V. (1994).\textit{\ J. Math. Phys.} \textbf{35}, 2878.

\bibitem{EfimovEtal1997}Efimov, G.V., von Waldenfels W. and Wehrse R. (1997) \textit{J. Quant. Spectrosc. Radiat. Transfer}, \textbf{58}, 355.

\bibitem{WehrseBaschek1999}Wehrse R. and Vaschek B. (1999),\textit{\ Physics Reports,} \textbf{311}, 187.

\bibitem{AnilePennisiSammartino1991}Anile, A. M. Pennisi, S. and Sammartino, M. (1991) \textit{J. Math.Phys.} \textbf{32}, 544.

\bibitem{HerreraJimenez1983}Herrera, L. and Jiménez, J. (1982),\textit{\ Phys. Rev. D}.,\textbf{\ 28,} 2987.

\bibitem{AguirreHernandezNunez1994}Aguirre, F., Hernández, H. and Núñez L.A. (1994), \textit{Astrophys and Space Sc. }\textbf{219, }153.

\bibitem{HerreraEtal1998}Herrera, L., Hernández, H., Núñez, L.A. and Percoco, U. (1998), \textit{Class. Quantum Grav}, \textbf{15}, 187. [Online article]: cited on 2 Oct 1997 \url{http://arXiv.org/abs/gr-qc/9710017}

\bibitem{Tolman1939}{Tolman, R.C. (1939) \textit{Phys. Rev.} \textbf{55}, 364.}

\bibitem{PatinoRago1983}{Patiño, A. and Rago, H. (1983) \textit{Lett. Nuovo Cimento} \textbf{38}, 321.}
\end{thebibliography}
\end{document}